\documentclass[tighten,floats,aps,preprintnumbers]{revtex4}
\usepackage{amsmath}
\usepackage{epsf}
\usepackage{color}

\begin{document}
\preprint{KUNS-2411,RIKEN-QHP-30}
\title{$\alpha$-cluster correlations and symmetry breaking 
in light nuclei}

\author{Yoshiko Kanada-En'yo}
\affiliation{Department of Physics, Kyoto University,
Kyoto 606-8502, Japan}

\author{Yoshimasa Hidaka}
\affiliation{Theoretical Research Division, Nishina Center, RIKEN, Wako 351-0198, Japan}

\begin{abstract}
$\alpha$-cluster correlations in the ground states of $^{12}$C and $^{16}$O are studied. 
Because of the $\alpha$ correlations, the intrinsic states of $^{12}$C and $^{16}$O have 
triangle and tetrahedral shapes, respectively. The deformations are regarded as 
spontaneous symmetry breaking of rotational invariance, and the resultant 
oscillating surface density is associated with 
a density wave (DW) state caused by the instability of Fermi surface with 
respect to a kind of $1p$-$1h$ correlations.
To discuss the symmetry breaking between uniform density states and the oscillating density state, 
a schematic model of a few clusters on a Fermi gas core in a one-dimensional finite box 
was introduced.  The model analysis suggests structure transitions from 
a Fermi gas state to a DW-like state via a BCS-like state, and to a Bose Einstein condensation (BEC)-like state 
depending on the cluster size relative to the box size. 
It was found that the oscillating density in the DW-like state
originates in Pauli blocking effects.
\end{abstract}

\maketitle

\section{Introduction}
Nuclear deformation is one of the typical
collective motions in nuclear systems.
It is known that ground states of nuclei often have static deformations 
in the intrinsic states, which are regarded as spontaneous symmetry breaking of the rotational invariance
due to collective correlations. Needless to say, the broken symmetry in the intrinsic states
is restored in nuclear energy levels, because total angular momenta are good quanta
in energy eigenstates of the finite system.
Not only normal deformations of axial symmetric quadrupole deformations 
but also triaxial, octupole, and super deformations have been attracting interests
in these decades. To investigate 
those deformation phenomena mean-field approaches have been applied, in particular, for heavy mass nuclei. 

In light nuclear systems, further exotic shapes due to cluster structures have been suggested.
For instance, a triangle shape in $^{12}$C and a tetrahedral one in $^{16}$O have been discussed 
based on the cluster picture that $^{12}$C and $^{16}$O are considered to be 
$3\alpha$ and $4\alpha$ systems. In old days, to understand spectra of $^{12}$C and $^{16}$O
non-microscopic $\alpha$-cluster models have been applied~\cite{wheeler37,dennison54}. 
From vibrations of the triangle structure of 
thee $\alpha$ particles and the tetrahedral one of four $\alpha$s, 
Wheeler has suggested the low-lying $J=3$ states in $^{12}$C and $^{16}$O~\cite{wheeler37}, 
which are now considered to correspond to 
the lowest negative-parity states $^{12}$C($3^-_1$, 9.64 MeV) and $^{16}$O($3^-_1$, 6.13 MeV) established experimentally. In 1970's, cluster structures of the ground and excited states in $^{12}$C and $^{16}$O 
have been investigated by using microscopic and semi-microscopic
cluster models~\cite{brink70,ikeda72,OCM,smirnov74,GCM,RGM,suzuki76,fujiwara80,bauhoff84},
a molecular orbital model~\cite{abe71}, and also 
a hybrid model of shell model and cluster model~\cite{takigawa71}.

For $^{12}$C, the ground state is considered to have the triangle deformation because of 
the $3\alpha$-cluster structure. In addition, a further prominent triangle $3\alpha$ structure has been 
suggested in $^{12}$C($3^-_1$, 9.64 MeV). 
Although the cluster structure of the ground state, $^{12}$C($0^+_1$), may not be so prominent as that of the
$^{12}$C($3^-_1$), the $J^\pi=0^+_1$ and $3^-_1$ states are often described by the rotation of 
the equilateral triangle $3\alpha$ configuration having the $D_{3\text{h}}$ symmetry.
In contrast to the geometric configuration suggested in $^{12}$C($0^+_1$) and $^{12}$C($3^-_1$),
a developed $3\alpha$-cluster structure with no geometric configuration has been suggested  
in the $0^+_2$ state assigned to $^{12}$C($0^+_2$, 7.66 MeV) by 
(semi-)microscopic three-body calculations of $\alpha$ clusters~\cite{OCM,GCM,RGM,fujiwara80}. 
In the $0^+_2$ state, three $\alpha$ particles are weakly interacting like a gas, for which
the normal concept of nuclear deformation may be no longer valid.
For the $0^+_2$ state, Tohsaki {\it et al.} has proposed a new interpretation of 
a dilute cluster gas state, where $\alpha$ particles behave as bosonic particles and 
condensate in the $S$ orbit~\cite{Tohsaki01}. 
This state is now attracting a great interest in relation with the 
Bose Einstein condensation (BEC) in nuclear matter~\cite{Ropke98}.

Let us consider the cluster phenomena in $^{12}$C from the viewpoint of symmetry breaking.
If the symmetry of the rotational invariance is not broken, a nucleus has a spherical shape. 
In the intrinsic state of $^{12}$C($0^+_1$), the spherical shape changes to the triangle shape 
via the oblate shape because of the $\alpha$-cluster correlation. 
It is the symmetry breaking from the rotational symmetry to the axial symmetry, and to the $D_{3\text{h}}$ symmetry.
In the group theory,  
it corresponds to 
 ${\rm O(3)} \rightarrow {\rm O(2)}\times Z_2 \rightarrow D_{3\text{h}}$, where 
the symmetry breaking from the continuous (rotational) group to the discrete (point) group occurs.
In the excited state, $^{12}$C($0^+_2$), the system again may have the continuous group O(3) symmetry.
It indicates that the cluster correlations in $^{12}$C($0^+_1$) and $^{12}$C($0^+_2$) may have
different features in terms of symmetry breaking. The triangle shape with the
$D_{3\text{h}}$ symmetry in $^{12}$C($0^+_1$) is characterized by the geometric configuration, 
while the $^{12}$C($0^+_2$) has no geometric configuration. 
Now a question arises: what is the mechanism of 
the symmetry breaking of the continuous group in  $^{12}$C($0^+_1$), which is restored 
again in $^{12}$C($0^+_2$). One of the key problems is the geometric configuration 
because of $\alpha$ correlations in the ground state. 
The triangle state has oscillating surface density along the oblate edge.
 It can be understood by 
the density wave (DW)-like correlation 
caused by the $1p$-$1h$ correlation carrying a finite momentum 
in analogy to the DW in infinite matter with inhomogeneous periodic densities, which has been 
an attractive subject in various field such as nuclear and hadron physics~\cite{overhauser60,brink73,llano79,ui81,tamagaki76,takatsuka78,migdal78,Dautry:1979bk,Deryagin:1992rw,Shuster:1999tn,Park:1999bz,Alford:2000ze,Nakano:2004cd,Giannakis:2004pf,Fukushima:2006su,Nickel:2009ke,Kojo:2009ha,Carignano:2010ac,Fukushima:2010bq}  as well as 
condensed matter physics~\cite{CDW,SDW}.
Indeed, in our previous work, we interpreted the triangle shape as
the edge density wave on the oblate state by extending the DW concept to 
surface density oscillation of finite systems~\cite{KanadaEn'yo:2011qf}. 
Then the structures of $^{12}$C($0^+_1$) and $^{12}$C($0^+_2$) 
may be associated with the DW and the BEC phases in infinite matter, respectively.
The mechanism of the geometric triangle shape in the finite system 
may give a clue to understand an origin of DW in infinite matter. 

Similar to $^{12}$C, a geometric configuration with a tetrahedral shape in $^{16}$O has been 
discussed in theoretical studies with cluster model calculations~\cite{wheeler37,brink70,bauhoff84} and
also with Hartree-Fock calculations~\cite{eichler70,onishi71,takami95}. 
The excited state, $^{16}$O($3^-_1$, 6.13 MeV), is understood by the tetrahedron 
vibration or the rotation of the tetrahedral deformation with the $T_d$ symmetry, while
the static tetrahedral shape in the ground state has not been confirmed yet.
The tetrahedron shape in $^{16}$O($0^+_1$) and $^{16}$O($3^-_1$) 
is supported in analysis of experimental data such as $E3$ transition strengths 
for $3^-_1 \rightarrow 0^+_1$~\cite{Robson:1979zz} 
and $\alpha$-transfer cross sections on $^{12}$C~\cite{elliott85}. 
The tetrahedral shape tends to be favored in cluster-model calculations~\cite{brink70,bauhoff84}.
However, 
Hartree-Fock calculations with tetrahedral deformed mean-field potentials
usually suggest the spherical $p$-shell closed state as the lowest solution for 
$^{16}$O except for the calculations using effective interactions with specially strong 
exchange forces~\cite{eichler70,onishi71,takami95}. 
If the ground state of $^{16}$O has the tetrahedral shape, 
it may suggest the breaking of the O(3) symmetry into the $T_d$ symmetry. 
In the excited $0^+$ states of $^{16}$O, 
$^{12}$C+$\alpha$ cluster structure was suggested in the $0^+_2$ state at 6.05 MeV
~\cite{ikeda72,suzuki76,fujiwara80,sheline60,horiuchi68}. Moreover, in analogy to the 3$\alpha$-cluster gas state of $^{12}$C($0^+_2$), 
a 4$\alpha$-cluster gas state with the $\alpha$ condensation feature has been suggested recently 
in a $0^+$ state above the 4$\alpha$ threshold~\cite{Funaki:2008gb,Funaki:2010px}.
Similarly to $^{12}$C, the possible tetrahedral shape in $^{16}$O may lead to 
symmetry breaking of the continuous group in  $^{16}$O($0^+_1$), which is restored 
in higher $0^+$ states. Again, the geometric configuration 
due to $\alpha$ correlations in the ground state should be 
one of the key problems. 

Our aim is to clarify the $\alpha$-cluster correlations with geometric configurations 
in the ground states of $^{12}$C and $^{16}$O, and understand the mechanism of the symmetry 
breaking from continuous (rotational) groups into discrete (point) groups.
We first confirm the problem whether the triangle and tetrahedron shapes
are favored in the intrinsic states of $^{12}$C and $^{16}$O. 
For this aim, we use a method of antisymmetrized molecular dynamics (AMD)~\cite{ENYOabc,ENYOsupp,AMDrev} and 
perform microscopic many-body calculations 
without assuming existence of any clusters nor geometric configurations. 
Variation after the spin-parity projections (VAP) is performed in the AMD framework~\cite{ENYO-c12}. 
The AMD+VAP method has been proved to be useful to describe
structures of light nuclei. With this method, one of the authors, Y. K-E., has 
succeeded to reproduce various properties of ground and excited states 
of $^{12}$C~\cite{ENYO-c12,KanadaEn'yo:2006ze}, and confirmed the formation and development of three
$\alpha$ clusters in $^{12}$C in the microscopic calculations with no cluster assumptions  
for the first time. The result was supported also by the work using the method of 
Fermionic molecular dynamics~\cite{Chernykh07}, which 
shows a similar method to the AMD.
 
In this paper, we apply the AMD+VAP method to $^{16}$O as well as $^{12}$C and analyze 
the intrinsic shapes of the ground states. 
We show that 
the geometric configurations having the approximate $D_{3\text{h}}$ and $T_d$ symmetry arise in the ground states
of $^{12}$C and $^{16}$O, respectively. 
To discuss appearance of the geometric configurations, we perform analysis using a simple
cluster wave functions of Brink-Bloch (BB) $\alpha$-cluster model~\cite{brink66},
while focusing on Pauli blocking effect on rotational motion of an $\alpha$ cluster. 
Important roles of the Pauli blocking effect 
in appearance of geometric configurations are described.
We also introduce a schematic model by considering clusters on a Fermi gas core 
in a one-dimensional (1D) finite box,
which can be linked with clusters at surface in a $3\alpha$ system. 
By analyzing the 1D-cluster wave function, in particular, looking at 
Pauli blocking effects from the core and those between
clusters, we try to conjecture what conditions favor BCS-like, DW-like, and BEC-like correlations.

This paper is organized as follows. In the next section, intrinsic shapes and cluster formation 
in the ground states of 
$^{12}$C and $^{16}$O are investigated based on the AMD+VAP calculation. In Sec.~\ref{sec:BB-cluster}
a Pauli blocking effect in $\alpha$-cluster systems and its role in $\alpha$-cluster correlations 
is described by analysis of BB $\alpha$-cluster wave functions. 
In Sec.~\ref{sec:1D-cluster}, the schematic model of clusters on the Fermi gas core 
in the 1D finite box is introduced and the roles of 
Pauli blocking effects in $\alpha$-cluster correlations are discussed. 
Summary and outlook are given in Sec.~\ref{sec:summary}.
The relations between $3\alpha$- and $4\alpha$-cluster wave functions
and triangle and tetrahedral deformed mean-field wave functions are explained in appendix
\ref{app:BB-MF}. In appendix \ref{sec:weak-coupling1} and \ref{sec:weak-coupling2}, 
features of weak-coupling wave functions in the 1D-cluster model are 
described. 

\section{Shapes and correlations in the ground states of $^{12}$C and $^{16}$O} \label{sec:AMD+VAP}

We discuss here intrinsic deformations of the ground states of $^{12}$C and $^{16}$O
based on the AMD+VAP calculation. 
The AMD method has been applied for various light mass nuclei and
has been successful in describing 
cluster structures as well as shell-model-like structures in light-mass nuclei. 
In the present work, the AMD+VAP method, i.e.,  variation after spin and parity projections  in the
AMD framework, is applied
to $^{12}$C and $^{16}$O. 
For the details of the framework, the reader is refereed to, for instance, Refs.~\cite{AMDrev,ENYO-c12}.  

\subsection{Variation after projection with AMD wave function}

In the AMD framework, we set a model space of wave functions and perform 
the energy variation to obtain the optimum solution in the model space.

An AMD wave function is given by a Slater determinant of Gaussian wave packets,
\begin{equation}
 \Phi_{\rm AMD}({\bf Z}) = \frac{1}{\sqrt{A!}} {\cal{A}} \{
  \varphi_1,\varphi_2,...,\varphi_A \},
\end{equation}
where the $i$th single-particle wave function is written by a product of
spatial ($\phi$), intrinsic spin ($\chi^\sigma$), and isospin wave functions ($\chi^\tau$) as
\begin{align}
 \varphi_i&= \phi_{{\bf X}_i} \chi^\sigma_i \chi^\tau_i,\\
 \phi_{{\bf X}_i}({\bf r}_j) & =   \left(\frac{2\nu}{\pi}\right)^{4/3}
\exp\bigl\{-\nu({\bf r}_j-\frac{{\bf X}_i}{\sqrt{\nu}})^2\bigr\},
\label{eq:spatial}\\
 \chi^\sigma_i &= (\frac{1}{2}+\xi_i)\chi_{\uparrow}
 + (\frac{1}{2}-\xi_i)\chi_{\downarrow}.
\end{align}
$\phi_{{\bf X}_i}$ and $\chi^\sigma_i$ are spatial and spin functions, and 
$\chi^\tau_i$ is the isospin
function fixed to be up (proton) or down (neutron). 
Accordingly, an AMD wave function
is expressed by a set of variational parameters, ${\bf Z}\equiv 
\{{\bf X}_1,{\bf X}_2,\cdots, {\bf X}_A,\xi_1,\xi_2,\cdots,\xi_A \}$.
The width parameter $\nu$ relates to the size parameter $b$ 
as $\nu=1/2b^2$ and it is chosen to be $\nu=0.19$ fm$^{-2}$ that minimizes energies of 
$^{12}$C and $^{16}$O.

The center positions ${\bf X}_1,{\bf X}_2,\cdots, {\bf X}_A$ of single-nucleon 
wave packets are independently 
treated as variational parameters. Thus existence of any clusters are not {\it a priori} assumed in the AMD framework. 
Despite of it, the model wave function can describe shell-model structures and 
cluster structures because of the antisymmetrizer and the flexibility of 
spatial configurations of Gaussian centers.
If a cluster structure is favored in a system, the corresponding cluster structure is 
automatically obtained in the energy variation. 

For even-even nuclei, the ground states are known to be $J^\pi=0^+$ states, i.e., they 
are symmetric for rotation.
Intrinsic deformation is understood as 
spontaneous symmetry breaking  with respect to the rotational invariance, which is restored in 
the $J^\pi=0^+$ ground states. It means that when an intrinsic state has a deformation 
the ground state is constructed by the spin and parity projections from the intrinsic state.
In more general, spin and parity are good quanta in energy eigenstates of nuclei
because of the invariance of the Hamiltonian for rotation and parity transformation. 
Therefore, to express a $J^\pi$ state, an AMD wave function is projected onto the spin-parity eigenstate, 
\begin{equation}
\Phi({\bf Z})=P^{J\pi}_{MK}\Phi_{\rm AMD}({\bf Z}),
\end{equation}
where $P^{J\pi}_{MK}$ is the spin-parity projection operator.

To obtain the wave function for a $J^\pi$ state, the energy variation is performed
for the spin-parity projected AMD wave function $\Phi({\bf Z})$ with respect to 
variational parameters $\{{\bf Z}\}$. This method is called variation after projection
(VAP). The AMD+VAP method has been applied to various light nuclei for structure study of
ground and excited states. 
For the ground states of $^{12}$C and $^{16}$O, 
we perform the variation of the energy expectation value 
$\langle \Phi({\bf Z})|H|\Phi({\bf Z}) \rangle/\langle \Phi({\bf Z})|\Phi({\bf Z}) \rangle$
for the $J^\pi=0^+$ projected wave function and
get the optimum parameter set $\{{\bf Z}\}$ that minimizes the
energy. Then, the AMD wave function $\Phi_{\rm AMD}({\bf Z})$ 
given by the optimized $\{{\bf Z}\}$ is regarded as the intrinsic wave functions of the ground state. 

An AMD wave function is expressed by a single Slater determinant; however, 
the spin-parity projected wave function is no longer a Slater determinant 
but it is a linear combination of Slater determinants except for the case that 
the AMD wave function before the projection is already a spin-parity $J^\pi$ eigenstate.
If the intrinsic state has a deformation
the projected wave function contains some kinds of correlations beyond Hartree-Fock 
approximation.

\subsection{Intrinsic structures of $^{12}$C and $^{16}$O}
We apply the AMD+VAP method to the ground states of $^{12}$C and $^{16}$O,
and discuss their intrinsic structures. 

\subsubsection{Density distribution}

\begin{figure}[th]
\centerline{\epsfxsize=7.5 cm\epsffile{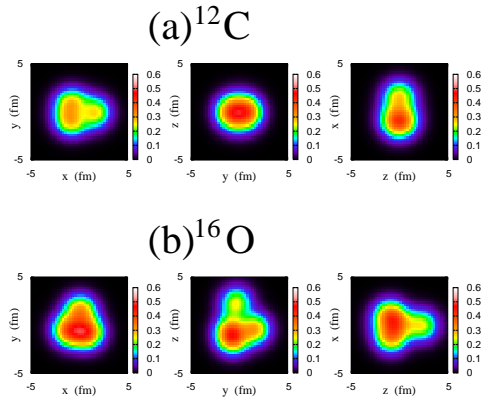}}
%\centerline{\epsfxsize=4 cm\epsffile{c12-dense-xy.eps}
%\epsfxsize=4 cm\epsffile{c12-dense-yz.eps}
%\epsfxsize=4 cm\epsffile{c12-dense-zx.eps}}
%\centerline{\epsfxsize=4 cm\epsffile{o16-dense-xy.eps}
%\epsfxsize=4 cm\epsffile{o16-dense-yz.eps}
%\epsfxsize=4 cm\epsffile{o16-dense-zx.eps}}
\caption{(color on-line) Density distributions in the intrinsic states for the ground states of 
(top) $^{12}$C  and (bottom) $^{16}$O  obtained by the AMD+VAP calculation. 
The density integrated on the $z$, $x$, and $y$ axes is plotted on the $x$-$y$, $y$-$z$, and $z$-$x$ planes. 
\label{fig:c12-o16.dense}}
\end{figure}

The ground states of $^{12}$C and $^{16}$O have the intrinsic deformations.  
The density distribution of the intrinsic wave functions $\Phi_{\rm AMD}$ are 
shown in Fig.~\ref{fig:c12-o16.dense}.
The result for $^{12}$C shows a triaxial deformation with a triangle feature, 
while $^{16}$O has a deformation with a tetrahedral feature.
The quadrupole deformation parameters $(\beta,\gamma)$ evaluated by the quadrupole moments
are $(\beta,\gamma)=(0.31,0.13)$ for $^{12}$C and  $(\beta,\gamma)=(0.25,0.09)$ for $^{16}$O.
The triangle and tetrahedral shapes are caused by $\alpha$-cluster correlations. Strictly speaking 
$\alpha$ clusters are not ideal $(0s)^4$ clusters but somewhat dissociated ones. Moreover, the triangle and tetrahedron
are not regular but distorted 
as an $\alpha$ cluster is situated slightly far from the other $\alpha$s.

\subsubsection{Oscillation of the surface density}
Let us consider the deformations of $^{12}$C and $^{16}$O 
from the viewpoint of symmetry breaking.
The highest symmetry is the sphere, which is realized in the closed $p^{3/2}$-shell and 
$p$-shell states for $^{12}$C and $^{16}$O, respectively. 
As a higher symmetry can break into lower symmetry due to multi-nucleon correlations,
the deformation mechanism of $^{12}$C is interpreted as follows.
Because of the $\alpha$-cluster correlations, 
the rotational symmetry in the sphere breaks to the axial symmetry in an oblate deformation 
and changes into the $D_{3\text{h}}$ symmetry in the regular triangle $3\alpha$
configuration, and then
it breaks to the distorted triangle. 
The symmetry change, spherical$\rightarrow$oblate$\rightarrow$triangle, corresponds to 
${\rm O(3)}\rightarrow {\rm O(2)}\times Z_2\rightarrow D_{3\text{h}}$. 
Similarly, the deformation of $^{16}$O is understood as
the rotational symmetry in the sphere breaks into the $T_d$ symmetry in the regular tetrahedral
$4\alpha$ configuration, and it breaks to the distorted tetrahedron.
The structure change from the spherical to the tetrahedron is the breaking of the O(3) symmetry to 
the $T_d$ symmetry. Note that the continuous (rotational) groups 
break to the discrete (point) groups in the triangle and tetrahedron deformations.

We discuss the connection of cluster correlations with the density wave, which is 
characterized by the static density oscillation at the surface.
The surface density oscillation occurs at the symmetry breaking from the axial symmetry (oblate shape) 
to the $D_{3\text{h}}$ symmetry (triangle) in $^{12}$C and 
that from the rotational symmetry (sphere) to the $T_d$ symmetry (tetrahedron) in $^{16}$O.
The triangle deformation contains the $(Y_3^{-3}-Y_3^{+3})/\sqrt{2}$ component with the 
$Y_2^0$ deformation in the density, while the tetrahedral one has 
the $(\sqrt{5}Y_3^{0}+\sqrt{2}Y_3^{-3}-\sqrt{2}Y_3^{+3})/3$ component, which can be transformed 
to $(Y_3^{-2}+Y_3^{+2})/\sqrt{2}$ by the rotation. Indeed, as described 
in appendix \ref{app:BB-MF}, 
an ideal $3\alpha$($4\alpha$)-cluster wave function with 
the triangle (tetrahedral) configuration has  
the density having the finite components of $(Y_3^{-3}-Y_3^{+3})/\sqrt{2}$ 
($(\sqrt{5}Y_3^{0}+\sqrt{2}Y_3^{-3}-\sqrt{2}Y_3^{+3})/3$) and it 
can be described by the DW-type particle-hole correlations in case of weak deformations.
Note that the DW in the triangle shape is characterized by
the particle-hole correlations on the Fermi surface carrying the finite angular momentum 
$(l,m)=(3,\pm 3)$ and that in the tetrahedral one is given by 
the particle-hole correlations with $(l,m)=(3,\pm 2)$ as described in the appendix.
Namely, the symmetry breaking is characterized by the 
$(Y_3^{-3}-Y_3^{+3})/\sqrt{2}$ component whose amplitudes
linearly relate to the order parameter of the DW correlations as shown
in Eqs.~(\ref{eq:3alpha-density}), (\ref{eq:4alpha-density}), and (\ref{eq:4alpha-density2}). 
In other words, the symmetry broken states have finite $(Y_3^{-3}-Y_3^{+3})/\sqrt{2}$ components in surface density
and show the oscillation density with the 
wave number three.

To analyze the surface density oscillation in the ground states of $^{12}$C and $^{16}$O, 
we perform the multipole decomposition of the intrinsic density obtained with the AMD+VAP calculation
\begin{equation}
\rho(r=R_0,\theta,\phi)=\bar\rho(R_0) \sum_{lm} \alpha_{lm} Y_l^m(\theta,\phi),
\end{equation}
at a certain radius $r=R_0$, and discuss the $(Y_3^{-3}-Y_3^{+3})/\sqrt{2}$ components. 
We take $R_0$ to be the root mean square radius of the intrinsic state.
$\bar\rho(R_0)$ is determined by the normalization $\alpha_{00}=1$, and
$\alpha_{lm}$ has the relation 
$\alpha_{lm}=(-1)^{m}\alpha^*_{l-m}$ because the density $\rho(r=R_0,\theta,\phi)$ is real.

\begin{figure}[th]
%\centerline{\epsfxsize=3.5 cm\epsffile{c12.dense-r.eps}\epsfxsize=4 cm\epsffile{c12.dense-phi.eps}}
\centerline{\epsfxsize=7 cm\epsffile{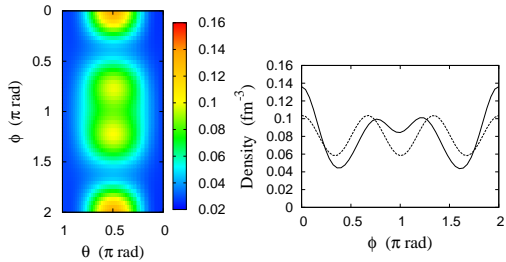}}
\caption{(color on-line) Left: density at $r=R_0$ for 
$^{12}$C calculated by the AMD+VAP is plotted on the $\theta$-$\phi$ plane.
Right: that at $r=R_0$ and $\theta=\pi/2$ line (the solid line). 
The density for the ideal $D_{3\text{h}}$ symmetry 
given only by the $(l,m)=(0,0)$ ,(2,0),(3,3) and (3,$-$3) components is also plotted (the dashed line).
$R_0$ is taken to be 2.53 fm.
}\label{fig:c12.theta-phi}
\end{figure}

\begin{figure}[th]
\centerline{\epsfxsize=6 cm\epsffile{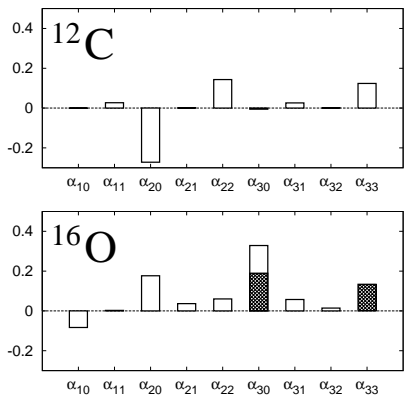}}
%\centerline{\epsfxsize=6 cm\epsffile{ylm-c12.eps}}
%\centerline{\epsfxsize=6 cm\epsffile{ylm-o16.eps}}
%\centerline{\epsfxsize=6 cm\epsffile{ylm-o16.l3n.eps}}
\caption{The coefficients $\alpha_{lm}$ of 
the multipole decomposition of the intrinsic density of (top) $^{12}$C and (bottom) $^{16}$O 
calculated by the AMD+VAP. The hatched areas in the bottom panel are the 
tetrahedron component $\sqrt{5}Y_3^{0}/3+\sqrt{2}Y_3^{+3}/3-\sqrt{2}Y_3^{-3}/3$
defined by $\alpha^{\rm hatch}_{30} \equiv \sqrt{5/2} \alpha_{33}$ and
$\alpha^{\rm hatch}_{33} \equiv \alpha_{33}$. The open area for the $Y_3^{0}$ component 
is defined by the relation $\alpha_{30}=\alpha^{\rm open}_{30}+\alpha^{\rm hatch}_{30}$.
\label{fig:ylm}}
\end{figure}

The density plot at $r=R_0$ on the $\theta$-$\phi$ plane for $^{12}$C  and that 
at $r=R_0$ and $\theta=\pi/2$ as a function of 
$\phi$ are shown in Fig.~\ref{fig:c12.theta-phi}. 
As seen clearly, the density on the oblate edge shows the oscillation 
with the approximate wave number three periodicity, which comes from the $\alpha$-cluster correlation.
In the right panel of Fig.~\ref{fig:c12.theta-phi}, we also plot the density for the ideal $D_{3\text{h}}$ symmetry 
given only by the $(l,m)=(0,0)$ ,(2,0),(3,3) and (3,$-$3) components
to show somewhat distortion of the AMD+VAP result from the ideal $D_{3\text{h}}$ symmetry.
In the coefficients $|\alpha_{lm}|$ shown in Fig.~\ref{fig:ylm}, it is found that the 
$Y_l^{\pm 3}$ components are actually finite indicating that 
the axial symmetry is broken to the triangle shape in $^{12}$C.

\begin{figure}[th]
\centerline{\epsfxsize=3.5 cm\epsffile{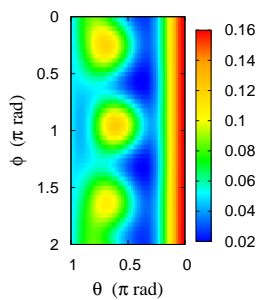}%\epsfxsize=3.5 cm\epsffile{o16.l3n.dense-r.eps}
}
\caption{ (color on-line) 
density at $r=R_0$ for 
$^{16}$C calculated by the AMD+VAP is plotted on the $\theta$-$\phi$ plane.
$R_0$ is chosen to be 2.81 fm. % for the $0^+_1$ state.%, and $R_0$=2.60 fm for the $3^-$ state.
\label{fig:o16.theta-phi}}
\end{figure}

For the density in $^{16}$O, the $\theta$-$\phi$ plot is shown in Fig.~\ref{fig:o16.theta-phi}, and 
the coefficients of the multipole decomposition are shown in Fig.~\ref{fig:ylm}. 
The tetrahedral component $\sqrt{5}Y_3^{0}/3+\sqrt{2}Y_3^{+3}/3-\sqrt{2}Y_3^{-3}/3$ 
is shown by the hatched boxes at $\alpha_{30}$ and $\alpha_{33}$ in Fig.~\ref{fig:ylm}. The open boxes 
indicate the distortion components from the tetrahedron. 
The distortion exists in the axial symmetry components, 
$\alpha_{30}$, $\alpha_{20}$, and $\alpha_{10}$, coming
from the spatial development of an $\alpha$ cluster from the others as explained before.

Thus, the 
intrinsic states of $^{12}$C and $^{16}$O show 
the surface density oscillation with the $(Y_3^{-3}-Y_3^{+3})/\sqrt{2}$ component.
The wave number three oscillation characterized by the $(Y_3^{-3}-Y_3^{+3})/\sqrt{2}$ component 
is understood by the $\alpha$-cluster correlations with triangle and tetrahedral deformations, which are 
interpreted as the DWs on the oblate and spherical shapes, i.e., the spontaneous symmetry breaking of
axial symmetry $\rightarrow$ $D_{3\text{h}}$ and rotational symmetry $\rightarrow$ $T_d$ symmetry.

\section{Pauli blocking effect in $\alpha$ correlations} \label{sec:BB-cluster}

As shown in the previous section, the
intrinsic states of $^{12}$C and $^{16}$O obtained with the AMD+VAP calculation
contain the $\alpha$-cluster correlations with the triangle and tetrahedral deformations. 
For $^{12}$C, the triangle deformation has been suggested in 
$3\alpha$-cluster models, in which $3\alpha$ state with a regular triangle configuration
is the energy minimum state~\cite{brink70,GCM,fujiwara80}. In contrast to such the geometric configuration as the triangle, 
the second $0^+$ state of $^{12}$C is considered to be a cluster gas state where 
3 $\alpha$ clusters are freely moving in dilute density 
like a gas without any geometric correlations between clusters~\cite{GCM,RGM,fujiwara80,Tohsaki01}.
The $0^+_2$ state is associated with the $\alpha$ condensation in analogy to BEC because it is regarded as a system of 
three $\alpha$s occupying the same $S$ orbit~\cite{Tohsaki01}.

From the viewpoint of symmetry breaking,
the symmetry is broken in the $0^+_1$ state with the triangle deformation and
it is restored in the $0^+_2$ state. 
In the $0^+_1$ state, the axial symmetry breaks down 
to the $D_{3\text{h}}$ symmetry due to the $3\alpha$-cluster structure. 
One of the characteristics of the $0^+_1$ state is the oscillating surface density caused by
the angular correlation of $\alpha$ clusters. Then the 
structure change from the $0^+_1$ to the $0^+_2$ is expected to connect with
the transition from the symmetry broken state with the angular correlation to the symmetric state with
no (or less) correlation between $\alpha$ clusters.

\begin{figure}[th]
\centerline{\epsfxsize=7.5 cm\epsffile{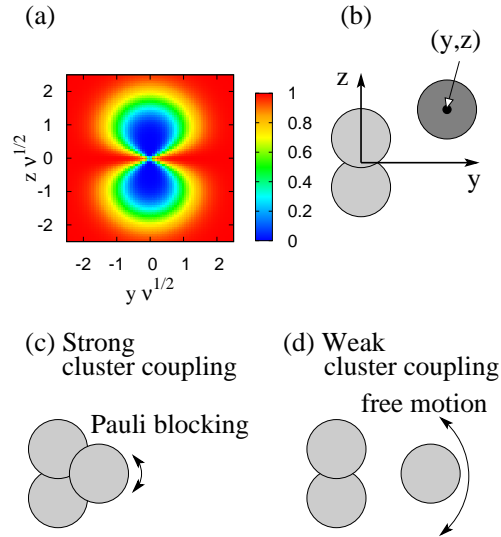}}
\caption{(a) The Pauli allowed and forbidden regions
for the rotation of the angle $\theta_y$ for the third $\alpha$ 
in the $(y,z)$-plane around the $2\alpha$ core. The norm 
$\tilde {\cal N}_{3\alpha}(y=r\cos \theta_y,z=r\sin \theta_y)$ normalized 
at $(y=0, z=r)$ is presented. The $\tilde {\cal N}_{3\alpha}\sim 1$ and $\tilde {\cal N}_{3\alpha}\sim 0$ regions
are the Pauli allowed and forbidden regions, respectively. 
(b) A schematic figure for the position of the third $\alpha$
around the $2\alpha$ core.
 (c) A schematic figure of the strong cluster coupling state for a compact $3\alpha$ state with the
strong the Pauli blocking effect, and 
(d) that of the weak cluster coupling state. See the text.
\label{fig:coupling}}
\end{figure}

The origin of the angular correlation in the $0^+_1$ state 
and the transition into the uncorrelated $0^+_2$  state can be 
understood by the Pauli blocking effect between clusters as follows.
Let us here consider motion of an $\alpha$ cluster around a $2\alpha$ core
in the BB 3$\alpha$-cluster model wave function. In the BB 3$\alpha$-cluster model, 
$\alpha$ clusters are located around certain positions ${\bf S}_1$, ${\bf S}_2$, and ${\bf S}_3$,
and the wave function is written as
\begin{equation}
\Phi_{BB}({\bf S}_1,{\bf S}_2,{\bf S}_3)=\frac{1}{\sqrt{A!}}{\cal A}\left\{ 
\Pi_{\tau\sigma}   \phi_{{\bf S}_1}{\cal X}_{\tau\sigma} \phi_{{\bf S}_2}{\cal X}_{\tau\sigma}
 \phi_{{\bf S}_3}{\cal X}_{\tau\sigma} \right\},
\end{equation}
where ${\cal X}_{\tau\sigma}$ is the spin-isospin wave function with 
$\tau=\{p,n\}$ and $\sigma=\{\uparrow,\downarrow\}$.
We assume that 2 $\alpha$ clusters placed at ${\bf S}_1=(0,0,d/2)$ 
and ${\bf S}_2=(0,0, -d/2)$ form the core.
%The parameter $d$ is fixed and taken to be a small value. 
The third $\alpha$ is placed 
at ${\bf S}_3=(0,y,z)$ for $y=r\cos \theta_y,z=r\sin \theta_y$ 
on the $(y,z)$-plane (see Fig.~\ref{fig:coupling}).
Because of Pauli blocking effect between clusters, 
the motion of the third $\alpha$ around the core is restricted in the Pauli allowed region. 
Particularly when the $\alpha$ exists near the core, rotational motion is strongly blocked because
of the existence of other $\alpha$ clusters. The Pauli allowed and forbidden areas 
for the rotation of the angle $\theta_y$ for the third $\alpha$ center 
in the $(y,z)$-plane are presented in Fig.~\ref{fig:coupling}.
In the figure, the norm ${\cal N}_{3\alpha}(y,z)=\langle \Phi_{BB}({\bf S}_1,{\bf S}_2,{\bf S}_3)|
\Phi_{BB}({\bf S}_1,{\bf S}_2,{\bf S}_3)\rangle$ of the BB wave function 
$\Phi_{BB}({\bf S}_1,{\bf S}_2,{\bf S}_3)$ with the parameters,
${\bf S}_1=(0,0,d/2)$, ${\bf S}_2=(0,0, -d/2)$, and ${\bf S}_3=(0,y,z)$ is shown in the $(y,z)$-plane.
$d$ is fixed and taken to be a small value. The norm is normalized by the value for the $\alpha$ 
on the y-axis,
\begin{equation}
\tilde {\cal N}_{3\alpha}(y,z)\equiv \frac{{\cal N}_{3\alpha}(y=r\cos \theta_y,z=r\sin \theta_y) }
{{\cal N}_{3\alpha}(y=r,z=0)}.
\end{equation}
The area with $\tilde {\cal N}_{3\alpha}\sim 1$ is the allowed region where 
the $\alpha$ feels no Pauli blocking with respect to the rotational motion, while the 
$\tilde {\cal N}_{3\alpha}\sim 0$ region is the blocked region where 
it feels the strong Pauli blocking effect from $\alpha$ clusters of the core. 
It means that, when the third $\alpha$ exists near the core, its angular motion is 
blocked by the $2\alpha$ core. As a result, the third $\alpha$ is confined in the 
Pauli allowed region around the $y$ axis, and it has the angular correlation against the
$2\alpha$ direction. 
Consequently, a compact $3\alpha$ state has a geometric structure of the 
triangle deformation and it has the surface density oscillation. 
On the other hand, as the cluster develops specially the Pauli blocking effect becomes weak. 
When the $\alpha$ is far enough from the core, it can freely move in the rotational mode. 
Then the angular correlation vanishes 
and the system transits to angular-uncorrelated state.  
We note that in cluster physics this transition is known to be 
the change between 
strong cluster coupling and weak cluster coupling states, where
the angular momentum of the inter-cluster motion couples with inner spins of clusters
strongly and weakly, respectively (see Fig.~\ref{fig:coupling}). What we call the cluster coupling 
is coupling between clusters and it is different from the terminology of strong and weak couplings 
in the BEC-BCS crossover phenomena, which relate to 
the coupling between nucleons in a pair (or in a cluster).

The $0^+_1$ state of $^{12}$C is considered to be the compact $3\alpha$ state in the 
strong Pauli blocking regime and corresponds to the angular correlated state 
with the surface density oscillation due to the triangle deformation.
In the $^{12}$C($0^+_2$), clusters spatially develop well and the system goes to the 
non-angular-correlation state.  
Strictly speaking, in the $^{12}$C($0^+_2$), all clusters develop 
to form a uncorrelating three $\alpha$ state associated with the $\alpha$ condensation,
where the radial motion is also important as well as the angular motion.
Nevertheless, we can say that, 
for the restoration of the broken symmetry with the surface density oscillation 
in the $0^+_1$ to the rotational symmetry in the $0^+_2$,
the transition from the angular correlated state to the uncorrelated state
in the cluster development is essential.

In the above discussion, we consider the angular motion in the intrinsic (body-fixed) frame, i.e., 
the $y$-$z$ plane. Since the system is symmetric for the rotation around the $z$-axis, the motion of the 
third $\alpha$ is free for the $z$-rotation. This rotational mode around the $z$ axis is nothing but 
the projection onto the $J_z=0$, which is usually performed 
in the $J^\pi=0^+$ projection of the intrinsic state.

Also in $^{16}$O, the angular motion of an $\alpha$ cluster at the surface 
is blocked by other three $\alpha$s. The Pauli blocking effect between 
$\alpha$ clusters causes the angular correlation of the tetrahedral configurations
in a compact $4\alpha$ state. For transition from the angular correlated cluster state of the ground state 
to the uncorrelated cluster state like a cluster gas, at least two $\alpha$ clusters need to spatially 
develop to move freely in the allowed region without feeling the Pauli blocking effect.
The $4\alpha$-cluster gas state in $^{16}$O has been suggested near the $4\alpha$ threshold energy. 
The suggested excitation energy $E_x\sim 15$ MeV is almost twice of $E_x=7.66$ MeV for 
the 3$\alpha$-cluster gas state in $^{12}$C. This might correspond to the energy cost of the 
spatial development for two $\alpha$s in $^{16}$O compared with 
that for an $\alpha$ in $^{12}$C.

\section{Clusters on a Fermi gas core in one dimensional box} \label{sec:1D-cluster}

\subsection{Concept of one-dimensional cluster model}
As mentioned, the triangle deformation of the $^{12}$C ground state can be interpreted as
the density wave on the axial symmetric oblate state and it is characterized by the 
static surface density oscillation with the wave number three on the edge of the oblate intrinsic state. 
The symmetry breaking in the ground state originates in 
the 4-body correlation in $\alpha$ clusters.  
The axial symmetry is broken by the angular correlation between $\alpha$ clusters 
due to the Pauli blocking effect. 
As $\alpha$ clusters develop, the Pauli blocking weakens and the angular correlation vanishes. 
Then the system transits from the symmetry broken state of the $^{12}$C($0^+_1$) to the 
symmetry restored state of the $^{12}$C($0^+_2$), where 
$\alpha$ clusters are freely moving in a dilute density like a gas.

In this section, we consider Pauli blocking effects
in a schematic model of two clusters on a Fermi gas core
in a one-dimensional (1D) finite box, 
and discuss how the translational invariance is broken to form 
an oscillating density (inhomogeneous) state and how the broken symmetry is restored
to a uniform density (homogeneous) state. 

Let us consider a schematic model for a simplified $3\alpha$ system consisting of 
two $\alpha$ clusters around the $\alpha$ core. To concentrate only on
angular correlations between two $\alpha$s, we 
ignore the radial coordinate degree of freedom
and assume that two $\alpha$ clusters are moving at a certain distance $r_0$ from the core. 
Taking the body-fixed frame so as one of two 
$\alpha$ clusters on the $z$-axis and the other $\alpha$ in the $y$-$z$ plane, then, 
the angular motion between two $\alpha$s can be reduced to 
the 1D problem, where the first cluster sits around the origin and the second cluster 
exists in a finite size $L=2\pi r_0$ box with the periodic boundary condition. 
For the core effect, we take into account only the Pauli blocking from the core particles, 
which is treated here by a Fermi gas core for simplicity.

We extend the 1D-cluster model and treat the cluster size $b$, 
the box size $L$, and the core Fermi momentum $k_c$ as parameters. 
In this model, we do not perform energy variation nor calculate energy eigenstates.
Moreover, the mechanism of cluster formation and the dynamical change of 
the cluster structure or the cluster size are beyond our scope. We give a model wave function with fixed parameters $b$,$L$,$k_c$
based on simple ansatz, and analyze Pauli blocking effect between a cluster and the 
core and that between two clusters in the given wave function. 
From behavior of the wave function and features of the Pauli blocking effects, we conjecture how the transition between
the Fermi gas, BCS-like, DW-like, and BEC-like states occurs. 

\subsection{Formulation of one-dimensional cluster model}
We explain details of the 1D-cluster model wave function.
Two $n$-body clusters are formed on a Fermi gas core
in a 1D box with the box size $L=2\pi r_0$. For an $\alpha$ cluster 
$n=4$ and a cluster consists of $p\uparrow$, $p\downarrow$, $n\uparrow$, and $n\downarrow$.
Basis single-particle wave functions in the periodic boundary condition are given as  
$e^{+ik x}/{\sqrt{2L}}$ and  $e^{-ikx}/{\sqrt{2L}}$, where the 
momentum $k=2\pi \tilde k/L$. Here $\tilde k$ is the dimensionless momentum $\tilde k=kr_0$ and 
takes $\tilde k=0,1,2,3,\cdots$. Core nucleons are assumed to form a Fermi surface at $k_c$. It means that 
$k\le k_c$ orbits are occupied by core nucleons and they are forbidden single-particle states for 
constituent nucleons of clusters. Using the dimensionless parameter $\tilde k_c=k_c r_0$,
the lowest allowed momentum is $k_F =\tilde k_F/r_0$ for $\tilde k_F\equiv \tilde k_c+1$ and
$k\ge k_F$ orbits are allowed. 

We assume the first cluster ($\alpha_1$) wave function that localized around $x=0$
in a Gaussian form, 
\begin{align} \label{eq:psi_1}
 \psi_{\alpha_1}&=\Pi_{\chi}\phi_0(x_{1\chi}){\cal X}_{\chi}, \\
 \chi&=\tau\sigma=\{p\uparrow, p\downarrow,n\uparrow,n\downarrow\},\\
 \phi_0(x)&=\frac{1}{\sqrt{2L}}\sum_{k \ge k_F} f(k)\left( e^{+ik x} + e^{-ikx}\right). 
\label{eq:spatial-alpha1}
\end{align}
$f(k)$ is the Fourier transformation of the Gaussian with the cluster size $b$ and given as 
\begin{equation}
f(k)=n_0 \exp(-\frac{b^2 k^2}{2}),
\end{equation}
where $n_0$ is determined by the condition,
\begin{equation}
\sum_{k \ge k_F} |f(k)|^2=1,
\end{equation}
to normalize $\phi_0(x)$ in the box. 
The wave function indicates that four species of nucleons, $p\uparrow, p\downarrow, n\uparrow, n\downarrow$, 
occupy the same spatial orbit with the Gaussian form forbidden partially by the Fermi gas core.
If $b/L$ and $k_F$ are small enough, the wave function is localized well around $x=0$.
Since there is no identical Fermion in a cluster, 
the antisymmetrizer can be omitted in the expression. 

Next we consider the second cluster ($\alpha_2$). 
Assuming that $\alpha_2$ is localized around a position $x=s$, the wave function is given 
by the shifted function, 
\begin{align}\label{eq:psi_2s}
 \psi^s_{\alpha_2}&=\prod_{\sigma}\phi_s(x_{2\chi}){\cal X}\chi,\\
 \phi_s(x)&=\frac{1}{\sqrt{2L}}\sum_{k \ge k_F} f(k)\left( e^{+ik (x-s)} + e^{-ik (x-s)}\right).
\end{align}
Then, the normalized wave function of the $2\alpha$ system with the parameter $s$ is written as
\begin{align}
 \Psi_s &= {\cal N}^{-2}_{\rm PB}(s)
\frac{1}{8!}{\cal A} \left\{ \psi_{\alpha_1} \psi^s_{\alpha_2} \right \} ,\\
{\cal N}_{\rm PB}(s)&\equiv|\frac{1}{\sqrt{2}}{\cal A}\{\phi_0\phi_s\}|^2 \\
&=\frac{1}{2}
|\phi_0(x_1)\phi_s(x_2)-\phi_0(x_2)\phi_s(x_1)|^2 \\
&= 1- \langle \phi_0\phi_s|\phi_s\phi_0\rangle.
\end{align}
${\cal N}_{\rm PB}(s)$ is the 
overlap norm for two identical nucleons. It is  
a function of the parameter $s$ for the localization center of the second cluster.
${\cal N}^4_{\rm PB}$ means the overlap norm for the $2\alpha$ system and it is an indicator
to evaluate the Pauli blocking effect between two clusters.
In the case that two clusters feel no Pauli blocking, ${\cal N}^4_{\rm PB}(s)=1$, while 
in the case that they feel complete Pauli blocking, ${\cal N}^4_{\rm PB}=0$.
It means that ${\cal N}^4_{\rm PB}(s)$ stands for the Pauli "allowedness" 
for $\alpha_2$ around $s$.

The density distribution for $\chi=\tau\sigma$ particles 
in the $2\alpha$ state is 
\begin{equation}
\rho^\chi_s(x)=\langle \Psi_s|\sum_{i \in \chi}\delta(x-x_i)| \Psi_s \rangle.
\end{equation}
Note that the density distributions for all kinds of nucleons are the same in the present cluster model 
and the total nuclear density is $\sum_\chi \rho^\chi_s(x)=4 \rho^\chi_s(x)$.
In a similar way, the $\chi=\tau\sigma$ density for the one-$\alpha$ state $\psi_{\alpha_1}$ is  
\begin{equation}
\rho^\chi_{\alpha_1}(x)=\langle \psi_{\alpha_1}|\sum_{i \in \chi }\delta(x-x_i)| \psi_{\alpha_1}\rangle.
\end{equation}

In the present model, we can express all the formulation with the dimensionless parameters 
$\tilde L=L/r_0=2\pi$, $\tilde b=b/r_0$, $\tilde s=s/r_0$, $\tilde x=x/r_0$, $\tilde k=kr_0$, 
$\tilde k_c=k_cr_0$, $\tilde k_F=k_F r_0$, and so on.
The dimensionless densities are also defined as $\tilde \rho=\rho r_0$.
Then, the state $|\Psi_s \rangle$ is specified with the dimensionless parameters $(\tilde k_F, \tilde b, \tilde s)$ 
and does not depend on the scaling factor $r_0$. 
Since the number of clusters is fixed to be two, 
a larger volume size $L$ corresponds to a lower cluster density. 
Because of the scaling, a larger volume size $L=2\pi r_0$ with a fixed cluster size $b$ 
is equivalent to a smaller cluster size $b$ with a fixed $L$.
That means, the parameter $\tilde b$ indicates the cluster size 
relative to the box size and also corresponds to the cluster density.
The $\tilde k_F=\tilde k_c+1$ is the lowest allowed momentum just above the core Fermi momentum $\tilde k_c$,
and it relates to the density of the core particles.

\subsection{Pauli blocking effect from the core to one cluster}
Let us describe the structure change of one cluster on the Fermi gas core
in the 1D-cluster model.
The coefficients $f(\tilde k)$ of the Fourier transformation and 
the density distribution $\rho^\chi_{\alpha_1}(\tilde x)$ for the $\alpha_1$ cluster
are shown in Figs.~\ref{fig:kf1} and \ref{fig:kf2} for $\tilde k_F=1$ and $\tilde k_F=2$, respectively.
Because of the Pauli blocking effect from the core as well as the finite volume effect, 
the structure of the cluster changes 
from the original Gaussian form 
depending of the parameters $\tilde b$ and $\tilde k_F$.
In the case of small $\tilde b$, which  corresponds to the small cluster size $b$ or the large volume size
$L$, the coefficient $f(\tilde k)$ is distributed widely and has a long 
tail toward the high momentum region, and the density is localized well 
around $\tilde x=0$ (and also $\tilde x=2\pi$ for the periodic boundary). 
With increase of $\tilde k_F$, the low momentum components are truncated 
and the density shows a oscillating tail. Nevertheless, if the cluster size   
$\tilde b$ is small enough, the density is still localized. With increase of the cluster size $\tilde b$, 
the density localization weakens and the density approaches the periodic one.
Then, the component of the lowest allowed orbit $\tilde k_F=\tilde k_c+1$ is dominant and 
the components of higher orbits decrease. The localization declines more rapidly 
in the case of the larger $\tilde k_F$. 
It means that, as the $\tilde b$ increases or  
as the $\tilde k_F$ increases, 
the spatial correlation between nucleons (inner correlation) in a cluster becomes 
weak. We call the case of the weak localization that the wave function has 
the dominant $\tilde k_F$ component and minor 
$\tilde k > \tilde k_F$  components ``the weak coupling regime'', 
and the opposite case of the strong localization with significant $\tilde k > \tilde k_F$ components
``the strong coupling regime''. In the weak coupling limit, the one-cluster density goes to 
$\cos^2(\tilde k_F \tilde x)/\pi$. 

\subsection{Pauli blocking between two clusters}

For two $\alpha$ clusters expressed by the cluster
wave function $\Psi_s$, the clusters $\alpha_1$ and $\alpha_2$ are assumed to be localized around 
$\tilde x=0$ and $\tilde x=\tilde s$, respectively. The Pauli blocking effect between $\alpha_1$ 
and $\alpha_2$ is evaluated by the overlap norm ${\cal N}^4_{\rm PB}(\tilde s)$, which is an indicator 
for Pauli allowedness. 
%In the wave function, 
The Pauli allowedness ${\cal N}^4_{\rm PB}(\tilde s)\sim 0$ means that 
the $\tilde s$ region is blocked by the $\alpha_1$ cluster and is a forbidden area for $\alpha_2$. 
The middle panels of Figs.~\ref{fig:kf1} and \ref{fig:kf2} show
the allowedness ${\cal N}^4_{\rm PB}(\tilde s)$ (thin solid lines) as well as ${\cal N}_{\rm PB}(\tilde s)$ (dashed lines) 
plotted as a function of $\tilde s$. In principle, the Pauli blocking effect reflects 
the probability of the $\alpha_1$ cluster, and therefore, the forbidden region corresponds to the 
relatively high density region of the $\alpha_1$ cluster. In case of a small cluster size $\tilde b$,
the allowed region exists widely and the forbidden region exists only 
in the small area close to $\tilde x=0$ and $\tilde x=2\pi$.
That is to say, the second cluster feels almost no Pauli blocking effect except for the region near 
the first cluster ($\alpha_1$).
With increase of $\tilde b$, the forbidden area spreads in a wide region around $\tilde x=0$ and $\tilde x=2\pi$, and the allowed region for $\alpha_2$ becomes narrow. In the weak coupling regime, where
the $\alpha_1$ density is periodic, 
the allowed region with ${\cal N}^4_{\rm PB}(\tilde s)\sim 1$
also shows the periodicity. 
Reflecting the $\cos^2(\tilde k_F x)$ periodicity of $\rho_{\alpha_1}(x)$, 
the areas $\tilde s\approx \pi 2m/2\tilde k_F$ ($m=0,\cdots,2 \tilde k_F$) are the forbidden regions, while  
the areas $\tilde s=\tilde s_j=\pi (2j-1)/2\tilde k_F$ ($j=1,\cdots,2 \tilde k_F$) are the allowed regions.

As mentioned, the parameter $\tilde b$ corresponds to the cluster size relative to the box size.
When the cluster density is low enough, the Pauli blocking effect is weak and 
almost all region is allowed for the $\alpha_2$ cluster except for the 
position close to the $\alpha_1$. On the other hand, in the case of the high cluster density
the Pauli blocking effect is strong and the allowed $s$ region is restricted 
in the periodic regions. 

\subsection{Transitions from strong coupling to weak coupling regimes} 

We first discuss the features of the two cluster wave function $\Psi_s$ with a 
fixed parameter $\tilde s$. It means that the center of the second cluster is located around 
the fixed position. Later, we will discuss 
how the spatial correlations between clusters (inter-cluster correlation) 
can be affected by the Pauli blocking. 

We choose $\tilde s_j=\pi (2j-1)/2\tilde k_F$ with $j=1$ and $j=\tilde k_F$, which are 
the allowed positions $\tilde s_j$ nearest to $\tilde x=0$ and $\tilde x=\pi$ and corresponds to the 
smallest and largest inter-cluster ($\alpha_1$-$\alpha2$) distances, respectively. 
The density distribution $\rho^\chi_s(\tilde x)$ in the 2$\alpha$-cluster wave function $\Psi_s$
are shown in the right panels of Figs.~\ref{fig:kf1} and \ref{fig:kf2} for $\tilde k_F=1$ and 
$\tilde k_F=2$.
In the strong coupling regime, for instance, the $(\tilde k_F,\tilde b)=(1.0, 0.25)$ state, 
the density shows the clear two peak structure and indicates that 
two clusters are well isolated without almost no overlap.
As the $\tilde b$ increases, the overlap region between clusters gradually increases and 
the density changes to the oscillation structure,
in particular, in the case of $j=\tilde k_F$.  
The density oscillation is remarkable, for instance, in the  $(\tilde k_F,\tilde b)=(1.0, 1.0)$ and
 $(\tilde k_F,\tilde b)=(2.0, 0.75)$ states, which shows the $2\tilde k_F+1$ periodicity.
With further increase of $\tilde b$, the density oscillation weakens and finally 
disappears to the uniform density, and the system goes to the Fermi gas limit  
with the Fermi momentum $\tilde k_F$.

In the present model, we put the $\alpha_2$ cluster around the fixed position $\tilde s$.
For more realistic wave functions of two $\alpha$ clusters, one should 
extend the model by taking into account the motion of the $\alpha_2$ cluster 
relative to the $\alpha_1$ cluster. 
Microscopically, it corresponds to superposing $\psi^s_{\alpha_2}$ with various values of the parameter 
$s$ as
\begin{equation}
\psi_{\alpha_2}=\int d\tilde s F(\tilde s) \psi^{s}_{\alpha_2}. 
\end{equation} 
The spatial correlation between clusters (inter-cluster correlation)
is expressed by the weight function $F(\tilde s)$. On the other hand, the correlation
between nucleons inside a cluster (inner correlation) is described by the intrinsic structure of a
single-cluster wave function of $\psi_{\alpha_1}$ or $\psi^{s}_{\alpha_2}$ determined by 
the parameters $\tilde b$ and $\tilde k_F$, and it is given by hand in the present model, where 
the cluster formation and its intrinsic structure are {\it a priori} assumed.

Moreover, the wave function should be projected to the 
total momentum $K_G=0$ for the center of mass motion (c.m.m.) of two clusters
so that the translational invariance is restored in the finite system.

In the following discussions, we do not treat the $\alpha_2$-cluster motion explicitly, 
but conjecture how the Pauli blocking may restrict the motion of the $\alpha_2$ cluster and affect 
to the spatial correlations between clusters (inter-cluster correlations). 
Since the area of high $\alpha_1$ density is blocked
as mentioned, the $\alpha_2$ cluster may move in the allowed regions with large
${\cal N}^4_{\rm PB}(\tilde s)$.
We assume that the interaction between clusters is weak and 
the Pauli blocking effect, which acts like an effective repulsion, gives the most important 
contribution 
in the relative motion between clusters. 

\subsubsection{BEC-like state}
Let us consider the strong coupling regime, where the cluster size $\tilde b$ is
small enough. The ${\cal N}^4_{\rm PB}(\tilde s)$ curve shows a wide open window of the allowed region
as in the case of $(\tilde k_F, \tilde b)=(1.0, 0.25)$. 
Since the Pauli blocking effect is weak, the 
$\alpha_2$ can move in the wide allowed region almost freely. 
It means
the strong inner correlation but almost no inter-cluster correlation.
In such the uncorrelated cluster state, two clusters may condensate 
approximately the zero momentum state in the ground state similarly to the BEC phase.

\subsubsection{DW-like state}
As the $\tilde b$ increases, the open window for the allowed region 
closes and $\alpha_2$ no longer can
move freely.
Instead, the allowed region becomes discrete, and 
the $\alpha_2$-cluster center may be confined 
around the allowed $\tilde s$ values, $\tilde s_j=\pi (2j-1)/2\tilde k_F$ ($j=1,\cdots,2 \tilde k_F$). 

For a possible wave function for $\alpha_2$, one may 
consider a superposition of wave functions with $\tilde s_j$ as 
\begin{equation}
\psi_{\alpha_2}=\sum_j F(\tilde s_j) \psi^{s_j}_{\alpha_2}. 
\end{equation} 
As seen in the density distribution $\rho^\chi_s(\tilde x)$ of two cluster states
shown in Figs.~\ref{fig:kf1} and ~\ref{fig:kf2}, 
the density oscillation shows the $2\tilde k_F+1$ periodicity for any $\tilde s_j$ values.
%It suggests that the density oscillations for different $\tilde s_j$
%are in phase and may coherently contribute
%to form a oscillating density even in the superposed wave function.

The density oscillation can be remarkable provided that 
the amplitude $F(\tilde s_j)$ for $j=\tilde k_F$ 
(nearest $\tilde s_j$ to the middle point $\tilde x=\pi$) is relatively larger  
than those for $\tilde s_j$ around $\tilde s=0$ and $2\pi$
because of the effective repulsion between clusters due to the Pauli blocking effect.
In other words, possible localization of the amplitude $F(\tilde s_j)$ 
because of the Pauli blocking effect causes 
the spatial correlation between clusters (inter-cluster correlation).
In such the case, the density oscillation shows
the clear $2\tilde k_F+1$ periodicity whose origin is the DW-type correlations. 
Indeed, the state contains dominantly 
the coherent $1p$-$1h$ components of a $\pm \tilde k_F+1$ particle and a $\pm \tilde k_F$ hole 
on the Fermi surface at $\tilde k_F$.
The $1p$-$1h$ correlation carries the $2\tilde k_F+1$ momentum 
and causes the $2\tilde k_F+1$ periodicity. This correlation is the similar to 
that of the DW phase. 
In the weak coupling approximation, the spatial wave function Eq.~(\ref{eq:spatial-alpha1}) for 
an $\alpha$ cluster
is approximated by the dominant 
$\tilde k_F$ component with a minor mixing of the $\tilde k_F+1$ component as
\begin{equation}
\phi_0(x)=\frac{1}{\sqrt{2L}} \bigl( \cos \tilde k_F \tilde x + \epsilon  \cos (\tilde k_F+1) \tilde x \bigr).
\end{equation}
In this approximation, it can be proved that the $2\alpha$ wave function $\Psi_s$ for $j=\tilde k_F$ actually
contains the dominant DW-type $1p$-$1h$ correlation 
in the particle-hole representation (See appendix \ref{app:weak-coupling}).

In the opposite case that there exists attractive force between clusters, 
the amplitudes $F(\tilde s_j)$ may gather to smaller $\tilde s_j$. 
It corresponds to the exciton(Exc)-type correlations described by 
the coherent $1p$-$1h$ components of a $\pm \tilde k_F+1$ particle and a $\mp \tilde k_F$ hole 
on the Fermi surface as described in appendix  \ref{app:weak-coupling}. 
In this case, the $1p$-$1h$ carries 
the momentum $|\tilde k_F+1\mp  \tilde k_F|=1$, which suggests that spatial density 
oscillation is not so remarkable. Indeed, such the feature is seen in the weaker density oscillation 
in $\rho^\chi_s({\tilde x})$ for $\tilde s=\tilde s_j$ ($j=1$) shown 
in Fig.~\ref{fig:kf2} (dashed lines in the right panels) for $\tilde k_F=2$.
We should comment that the system for $k_F=1$ is a special case that 
the small $\tilde s$ region, for instance, the region $\tilde s < \pi/4$ is forbidden 
and the Exc-type correlations are suppressed.   

\subsubsection{BCS-like state}
With further increase of $\tilde b$, the $\tilde k_F$ component becomes more dominant.
In the case $f(\tilde k) \ll f(\tilde k_F)$ for $\tilde k > \tilde k_F$, 
the structure of the Pauli allowed area for $\alpha_2$ approaches 
the pure periodic one 
following the periodicity of the $\alpha_1$ density 
$\rho_{\alpha_1}(\tilde x)\approx \cos^2(\tilde k_F \tilde x)/\sqrt{\pi}$. 
Then, a linear combination of $\psi^{s_j}_{\alpha_2}$ with an equal weight may be the lowest state
to restore the translational invariance because $\Psi_{s}$ for different $\tilde s_j$ may 
degenerate energetically.
It means that the state no longer has the spatial correlation between clusters (inter-cluster correlation)
and it corresponds to a BCS-like state. Namely, the BCS-like state has the weak inner correlation 
and no inter-cluster correlation.

As described in appendix \ref{app:weak-coupling}, 
in the weak coupling approximation, 
the total c.m.m. momentum $K_G=0$ state projected from 
the equal weight linear combination of 
two cluster wave functions $\Psi_{s}$ is equivalent to the BCS-like state containing 
$2p$-$2h$ configurations of a 
$(\tilde k_F+1,-\tilde k_F-1)$ $\chi_\alpha\chi_\beta$ particle pair and a $(\tilde k_F,-\tilde k_F)$ $\chi_\alpha\chi_\beta$ hole pair
[see Eq.~(\ref{eq:BCS})].
In the $2p$-$2h$ state, all kinds of pairing is coherently mixed so as to keep 
the spin-isospin symmetry of $\alpha$ clusters.

\subsubsection{Fermi-gas state and correlations}
In the large $\tilde b$ limit, excited components of $\tilde k > \tilde k_F$ 
vanish and the system goes to the Fermi gas (FG) state with the Fermi surface at 
$\tilde k_F$. 
Needless to say, the FG state has no inner correlation nor
inter-cluster correlation.
This is nothing but the uncorrelated state. On the other hand, 
the correlated states are characterized by configurations of 
excited $\tilde k > \tilde k_F$ components.
In the weak coupling regime, 
we can recognize DW-like, Exc-like, BCS-like states, or mixing of them
by correlations in $1p$-$1h$ and $2p$-$2h$ configurations on the Fermi surface. 

\subsubsection{Diagram of structure transitions}
As discussed above, in the present schematic model for two clusters on the Fermi gas core in the 1D box,  
the cluster state is expected to show the BEC-like, the DW/Exc-like, BCS-like, or FG behaviors
depending on the cluster size $\tilde b$ and the lowest allowed momentum $\tilde k_F$.
We here assume the criterion to judge a correlation type for a wave function with 
given $\tilde b$ and $\tilde k_F$ values, and 
show a diagram of structure transitions on the $\tilde b$-$\tilde k_F$ plane.
  
For the criterion, we define 
$\Delta \tilde k$ by the deviation of $\tilde k$ from the lowest allowed momentum $\tilde k_F$.
\begin{equation}
(\Delta \tilde k)^2= \sum_{\tilde k\ge \tilde k_F} f^2(\tilde k) (\tilde k-\tilde k_F)^2.
\end{equation}
It indicates the deviation from the FG state $|HF\rangle$.
In the case $\Delta \tilde k \ll 1$, $\tilde k \ge \tilde k_F+2$ components are negligible 
and the coefficient $\epsilon\equiv f(\tilde k_F+1)$ of the $\tilde k_F+1$ component approximates  $\epsilon \approx \Delta \tilde k$. 
In the $\Delta \tilde k=0$ limit, the system goes to the FG state. 
For the criterion to classify a wave function with given $\tilde b$ and $\tilde k_F$
to a correlation type, we use 
the deviation $\Delta \tilde k$ and the Pauli allowedness ${\cal N}^4_{PB}(\tilde s)$
at the middle point of the box $\tilde s= \tilde L/2 =\pi$.

In Table \ref{tab:diagram}, we list the criterion for various correlation types. 
In the table, we also show the typical examples of the $\tilde b$ values for $\tilde k_F=1$ 
and $\tilde k_F=2$ that 
satisfy the criterion. The densities and the Pauli allowedness for the corresponding states are already
shown in Figs.~\ref{fig:kf1} and \ref{fig:kf2}. 
The BEC-like state is expected to appear when the Pauli allowed region is widely open. 
Then, we adopt the condition ${\cal N}^4_{PB}(\tilde s =\pi)>0.8$ for the BEC-like state.
The DW-like and/or Exc-like states may appear when the Pauli allowed region is restricted. 
For this condition, we use ${\cal N}^4_{PB}(\tilde s = \pi) <0.1$. 
The DW/Exc-like states may change to the BCS-like state
when the $\alpha_1$-cluster density $\rho_{\alpha_1}(x)$ approaches 
 $\cos^2(\tilde k_F x)/\sqrt{\pi}$ and the Pauli allowed area for $\alpha_2$ becomes
almost periodic. For the criterion that $\rho_{\alpha_1}(\tilde x) \approx
\cos^2(\tilde k_F \tilde x)/\sqrt{\pi}$, we adopt the ratio of 
the $\alpha_1$ density at $\tilde x=0$ to that at $\tilde x=\tilde L/2=\pi$.  
In the weak coupling regime of $\Delta \tilde k \ll 1$, 
the ratio $\rho^\chi_{\alpha_1}(\tilde x=0)/\rho^\chi_{\alpha_1}(\tilde x=\pi)$ 
is approximately given by 
$1-4 \epsilon \approx 1-4\Delta \tilde k$, therefore, we use $\Delta \tilde k$ for the measure.
We apply $\Delta \tilde k>0.05$ for the  DW/Exc-like states. This approximately
corresponds to the ratio $\rho^\chi_{\alpha_1}(\tilde x=0)/\rho^\chi_{\alpha_1}(\tilde x=\pi)< 0.8$. 
With the decrease of $\Delta \tilde k$, the $2\alpha$ wave function may gradually change to the FG state 
via the BCS-like state. 
Since the higher momentum components $\tilde k> \tilde k_F$
nearly equals to $(\Delta \tilde k)^2$ in the weak coupling regime, we use  $\Delta \tilde k$ as the measure
for structure transitions from the DW/Exc-like to the FG state as listed in the table.
For instance, the condition $\Delta \tilde k < 0.001$ for the FG state indicates
the contamination of  $\tilde k> \tilde k_F$ components is less than $10^{-6}$.

\begin{table}[htb]
\caption{\label{tab:diagram} 
For the criterion to classify the 1D-cluster wave function with given $\tilde b$ and $\tilde k_F$
into various types of correlation.
The deviation $\Delta k$ and the Pauli allowedness ${\cal N}^4_{PB}(\tilde s = \tilde L/2 =\pi)$ are
used for the criterion.
The typical examples of the $\tilde b$ values for $\tilde k_F=1$ and $\tilde k_F=2$ that 
satisfy the criterion are also shown in the table.
}
\begin{center}
\begin{tabular}{cccc}
\hline
correlation & criterion & \multicolumn{2}{c}{example} \\
& & $\tilde k_F=1$ & $\tilde k_F=2$ \\
FG & $\Delta \tilde k < 0.001$ &  &  $\tilde b=2.0$ \\
FG-BCS crossover &  $ 0.001 \le \Delta \tilde k < 0.005$ & $\tilde b=2.0$ &  $\tilde b=1.5$ \\
BCS-like &  $ 0.005 \le \Delta \tilde k < 0.01$ & & \\
BCS-DW/Exc crossover & $ 0.01 \le \Delta \tilde k < 0.05$ &  $\tilde b=1.5$ &  $\tilde b=1.25$\\
DW/Exc-like &  $ 0.05 \le \Delta \tilde k$ and ${\cal N}^4_{PB}(\tilde L/2) <0.1$ & 
$\tilde b=1.0$ &  $\tilde b=0.75$\\ 
DW/Exc-BEC crossover & $0.1 \le {\cal N}^4_{PB}(\tilde L/2) < 0.8$ & $\tilde b=0.5$ &  $\tilde b=0.5$\\
BEC-like & $0.8 < {\cal N}*^4_{PB}(\tilde L/2)$ & $\tilde b=0.25$ &  $\tilde b=0.25$ \\
\hline
\end{tabular}
\end{center}
\end{table}

\begin{figure}[th]
\centerline{\epsfxsize=7 cm\epsffile{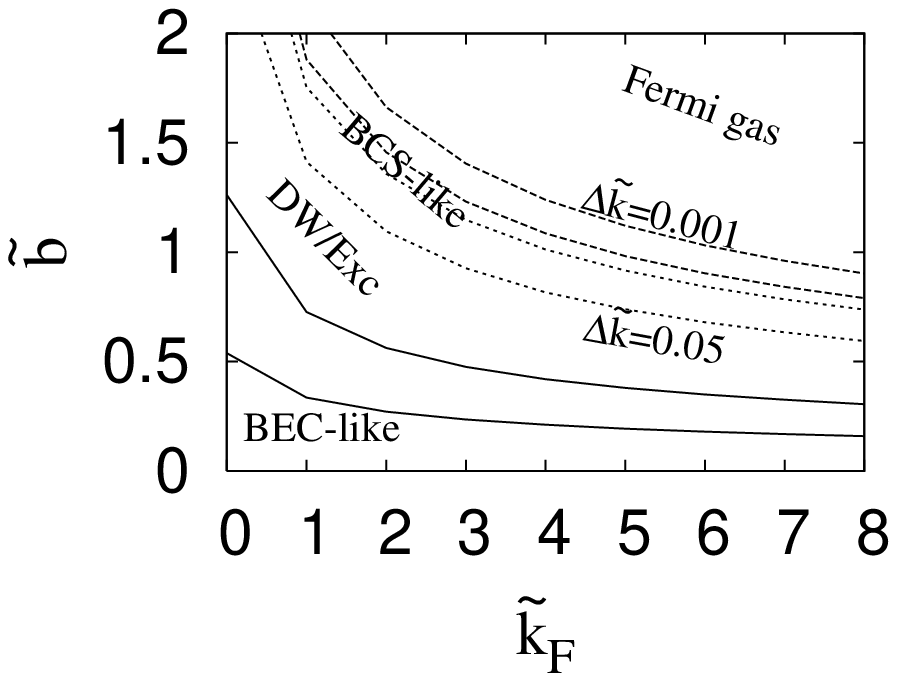}
}
\caption{Diagram of structure transitions between the Fermi gas state, BCS-like state, DW/Exc-like state, and
BEC-like state in the schematic 1D-cluster model. The criterion for the boundary are listed in
\ref{tab:diagram}. The conditions of ${\cal N}^4_{PB}(\tilde L/2) = 0.1$ and $0.8$ are shown by the solid lines, 
while the conditions of $\Delta \tilde k=0.001, 0.005, 0.01, 0.05$ are shown by the dashed lines. 
}\label{fig:phase}
\end{figure}

The diagram of the structure transitions between 
the FG state, BCS-like state, DW/Exc-like state, and
BEC-like state on the $\tilde k_F$-$\tilde b$ plane is shown in Fig.~\ref{fig:phase}.

For a system with higher $\tilde k_F$,  nucleons in a cluster couple more weakly to each other 
because of Pauli blocking from core nucleons, and therefore
the FG region is wider in the diagram.
However, one should care about that the assumption of the sharp surface for the core 
Fermi momentum might be 
inadequate, in particular, in case of high $\tilde k_F$. The core Fermi surface may diffuses 
in correlated states. If the surface diffuses, the lower orbitals below $\tilde k_F$ are  
partially allowed for nucleons in clusters. 
Then, the weakening of the cluster localization by the Pauli blocking from core nucleons 
can be quenched. 
The present model should be extended by incorporating the surface diffuseness of 
the core Fermi surface.

We also should comment that, in a real system, 
two parameters $\tilde b$ and $\tilde k_F$ are uncontrollable in general.
As explained before, $\tilde b$ indicates the cluster size relative to the box size (the cluster density), 
while $\tilde k_F$ relates to the core density. The size of clusters on the Fermi gas core
should be determined dynamically as a consequence of nuclear interactions between constituent nucleons
in clusters.
It should be also affected by the neighboring cluster, i.e., cluster density as well as $\tilde k_F$.  
Moreover, the ratio of cluster nucleons to the core nucleons should be determined dynamically.
The extension of the present model by taking into account
dynamical change of cluster structure with the use of effective nuclear interactions is an 
important issue to be solved in future study. 

\subsection{Correspondence to finite nuclei} 
\begin{figure}[th]
\centerline{\epsfxsize=10 cm\epsffile{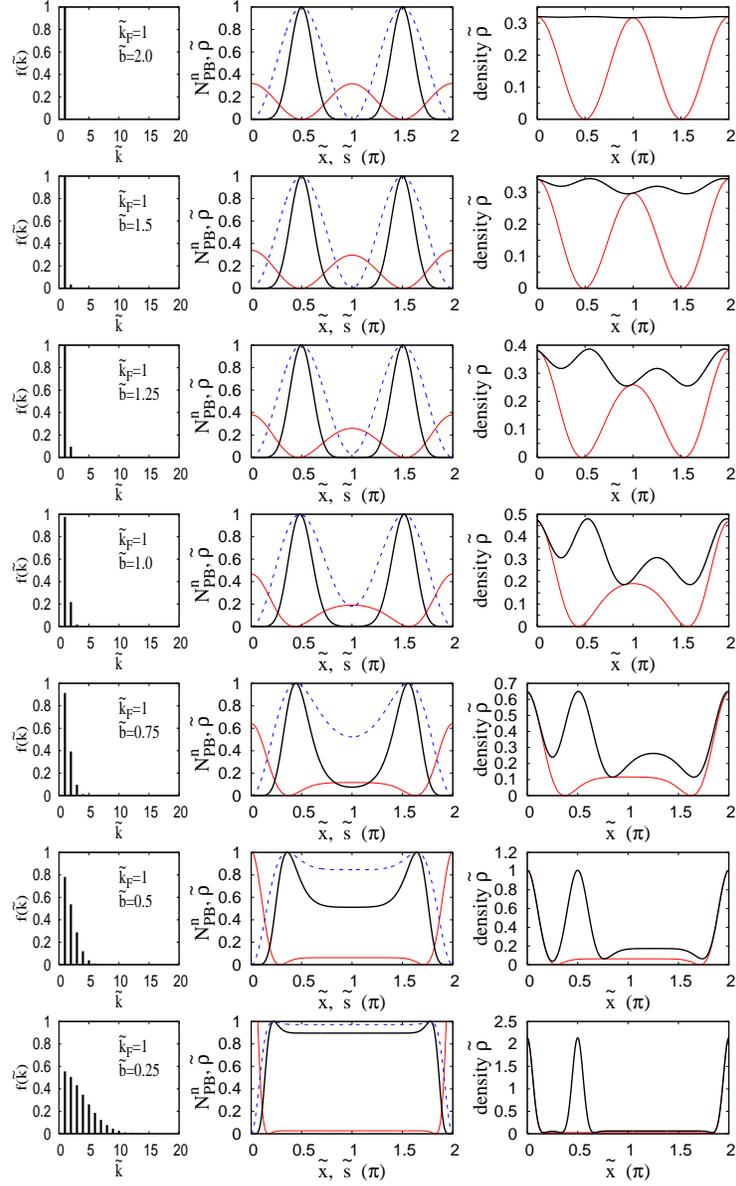}}
%\centerline{\epsfxsize=3.7 cm\epsffile{k0-b2.0-k-p.eps}\epsfxsize=4.5 cm\epsffile{k0-b2.0-rho1-p.eps}\epsfxsize=4.5 cm\epsffile{k0-b2.0-rho2-p.eps}}
%\centerline{\epsfxsize=3.7 cm\epsffile{k0-b1.5-k-p.eps}\epsfxsize=4.5 cm\epsffile{k0-b1.5-rho1-p.eps}\epsfxsize=4.5 cm\epsffile{k0-b1.5-rho2-p.eps}}
%\centerline{\epsfxsize=3.7 cm\epsffile{k0-b1.25-k-p.eps}\epsfxsize=4.5 cm\epsffile{k0-b1.25-rho1-p.eps}\epsfxsize=4.5 cm\epsffile{k0-b1.25-rho2-p.eps}}
%\centerline{\epsfxsize=3.7 cm\epsffile{k0-b1.0-k-p.eps}\epsfxsize=4.5 cm\epsffile{k0-b1-rho1-p.eps}\epsfxsize=4.5 cm\epsffile{k0-b1-rho2-p.eps}}
%\centerline{\epsfxsize=3.7 cm\epsffile{k0-b0.75-k-p.eps}\epsfxsize=4.5 cm\epsffile{k0-b0.75-rho1-p.eps}\epsfxsize=4.5 cm\epsffile{k0-b0.75-rho2-p.eps}}
%\centerline{\epsfxsize=3.7 cm\epsffile{k0-b0.5-k-p.eps}\epsfxsize=4.5 cm\epsffile{k0-b0.5-rho1-p.eps}\epsfxsize=4.5 cm\epsffile{k0-b0.5-rho2-p.eps}}
%\centerline{\epsfxsize=3.7 cm\epsffile{k0-b0.25-k-p.eps}\epsfxsize=4.5 cm\epsffile{k0-b0.25-rho1-p.eps}\epsfxsize=4.5 cm\epsffile{k0-b0.25-rho2-p.eps}}

\caption{(Color on-line) The results of one and two clusters on the Fermi gas core in a 
one-dimension box.
Left: The coefficients $f(\tilde k)$ of the Fourier transformation.
Middle: The density $\rho^\chi_{\alpha_1}(\tilde x)$ 
of the first $\alpha$ located around $\tilde x=0$ (red thin lines), 
the PB effect ${\cal N}^4_{PB}(\tilde s)$ for the second $\alpha$ (black solid lines).
${\cal N}_{PB}(\tilde s)$ is also shown (blue dashed lines).
Right: The density $\rho^\chi_{s}(\tilde x)$ (black solid lines). The $\tilde s$ for the center 
position of $\alpha_2$ is chosen to be $\tilde s_j$ with $j=\tilde k_F$, which is the closest
$\tilde s_j$ to $\tilde x=\tilde L/2=\pi$ among the allowed $\tilde s$ values. 
The density $\rho^\chi_{\alpha_1}(\tilde x)$ is also shown for comparison(red thin lines).
The results for the cluster size 
$\tilde b=2.0$, 1.5, 1.25, 1.0, 0.75, 0.5, 0.25 are shown. }
\label{fig:kf1}
\end{figure}

\begin{figure}[th]
\centerline{\epsfxsize=10 cm\epsffile{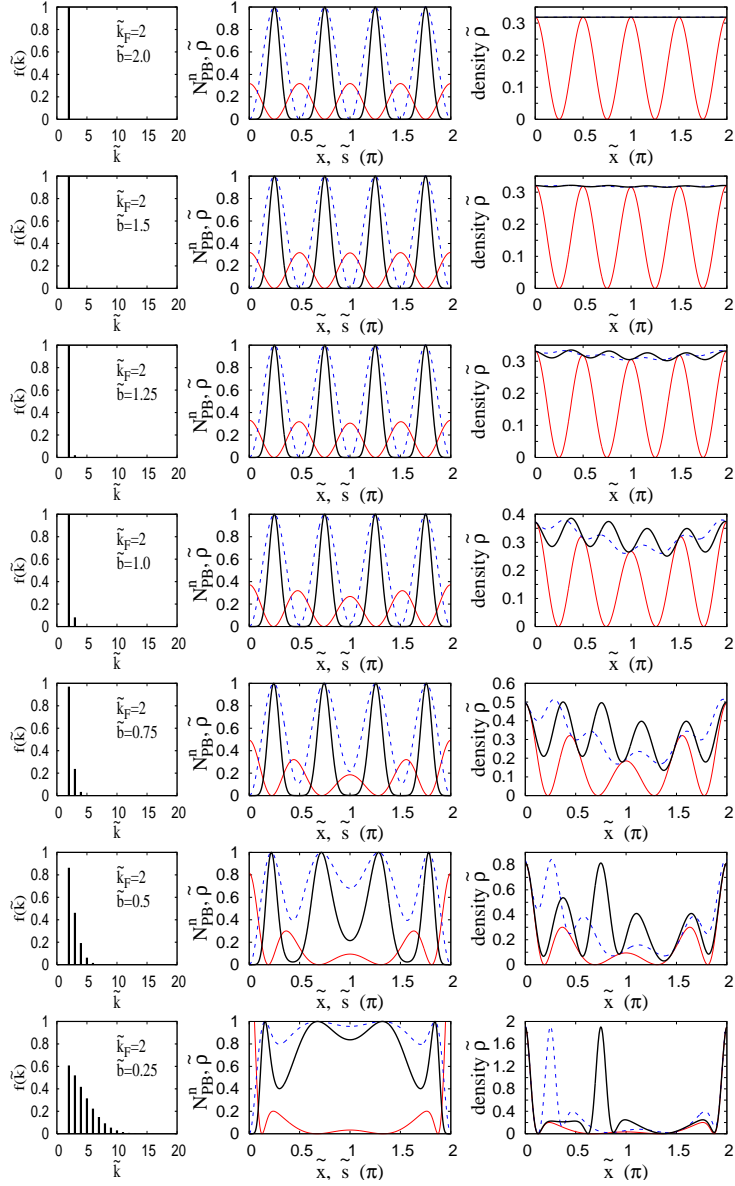}}
%\centerline{\epsfxsize=3.7 cm\epsffile{k1-b2.0-k-p.eps}\epsfxsize=4.5 cm\epsffile{k1-b2.0-rho1-p.eps}\epsfxsize=4.5 cm\epsffile{k1-b2.0-rho2-p.eps}}
%\centerline{\epsfxsize=3.7 cm\epsffile{k1-b1.5-k-p.eps}\epsfxsize=4.5 cm\epsffile{k1-b1.5-rho1-p.eps}\epsfxsize=4.5 cm\epsffile{k1-b1.5-rho2-p.eps}}
%\centerline{\epsfxsize=3.7 cm\epsffile{k1-b1.25-k-p.eps}\epsfxsize=4.5 cm\epsffile{k1-b1.25-rho1-p.eps}\epsfxsize=4.5 cm\epsffile{k1-b1.25-rho2-p.eps}}
%\centerline{\epsfxsize=3.7 cm\epsffile{k1-b1.0-k-p.eps}\epsfxsize=4.5 cm\epsffile{k1-b1.0-rho1-p.eps}\epsfxsize=4.5 cm\epsffile{k1-b1.0-rho2-p.eps}}
%\centerline{\epsfxsize=3.7 cm\epsffile{k1-b0.75-k-p.eps}\epsfxsize=4.5 cm\epsffile{k1-b0.75-rho1-p.eps}\epsfxsize=4.5 cm\epsffile{k1-b0.75-rho2-p.eps}}
%\centerline{\epsfxsize=3.7 cm\epsffile{k1-b0.5-k-p.eps}\epsfxsize=4.5 cm\epsffile{k1-b0.5-rho1-p.eps}\epsfxsize=4.5 cm\epsffile{k1-b0.5-rho2-p.eps}}
%\centerline{\epsfxsize=3.7 cm\epsffile{k1-b0.25-k-p.eps}\epsfxsize=4.5 cm\epsffile{k1-b0.25-rho1-p.eps}\epsfxsize=4.5 cm\epsffile{k1-b0.25-rho2-p.eps}}

\caption{Same as Fig.~\ref{fig:kf1} but for $\tilde k_F=2$. 
The blue dashed lines indicate the 
the density $\rho^\chi_{s}(\tilde x)$ for the smallest allowed $\tilde s$ value, 
$\tilde s_j$ with $j=1$. 
The results for the cluster size 
$\tilde b=2.0$, 1.5, 1.25, 1.0, 0.75, 0.5, 0.25 are shown. 
}
\label{fig:kf2}
\end{figure}

As shown in the AMD+VAP calculation, 
three  $\alpha$ clusters are formed in the ground state of $^{12}$C 
even though the existence of any clusters is not {\it a priori} assumed in the framework.
Once three $\alpha$ clusters are formed, it can be associated with 
the schematic 1D-cluster model.
Angular correlation between two $\alpha$ clusters around the $\alpha$ core  
corresponds to the spatial correlation between two $\alpha$ clusters 
in the 1D-cluster model with $\tilde k_F=1$. 
The cluster size $b$ is 1.62 fm for $\nu=0.19$ fm$^{-2}$ used in the AMD calculations.
If we adopt the r.m.s. matter radius $r_\alpha=1.72$ fm of the core $\alpha$
as the radial size $r_0$, the dimensionless $\tilde b=b/r_0$ is estimated to be $\tilde b=0.94$.
Or if we use the r.m.s. radius of cluster positions of 3 $\alpha$s
evaluated from $r^2_0+r^2_\alpha=R_0^2$, we get $\tilde b=0.87$. In both cases, $\tilde b\sim 1$.
As already described,
the state with $(\tilde k_F,\tilde b)=(1.0,1.0)$ in the 1D-cluster model
shows the remarkable density oscillation with the wave number three in the DW-like regime.
It is consistent with the intrinsic structure with the 
$^{12}$C($0^+_1$) obtained with the AMD+VAP calculation.
Indeed, the $^{12}$C($0^+_1$) has the $(Y_3^{-3}-Y_3^{+3})/\sqrt{2}$ component
and the surface density shown in Fig.~\ref{fig:c12.theta-phi}
is similar to the oscillating density in the 1D-cluster state for $(\tilde k_F,\tilde b)=(1.0,1.0)$ 
in Fig.~\ref{fig:kf1}.

As discussed in Ref.~\cite{KanadaEn'yo:2011qf}, the $\alpha$ correlation in the pentagon 
ground state of $^{28}$Si can be interpreted as the density wave on the $sd$-shell oblate state. 
The $^{28}$Si ground state is associated with the 1D-cluster model with $\tilde k_F=2$
considering a core consisting of the spherical $^{16}$O and four nucleons in oblate orbits. 
Then, the pentagon shape can be understood again by the DW-like state in the 1D-cluster model 
wave function with the $2\tilde k_F+1=5$ periodicity. 

In case of the $\alpha$ correlation in the $^{16}$O ground state, the tetrahedron shape can not
be connected directly to the 1D problem. However, when we focus only on the 
$(Y_3^{-3}-Y_3^{+3})/\sqrt{2}$ component of the intrinsic density in the $^{16}$O ground state, 
the density oscillation is characterized by the wave number three periodicity similar to 
that of the triangle shape in $^{12}$C($0^+_1$) and the deformation feature is associated with 
the DW-type correlation in the 1D cluster model as in $^{12}$C.

\section{Summary and outlook}\label{sec:summary}

We investigated 
$\alpha$-cluster correlations in the ground states of $^{12}$C and $^{16}$O while 
focusing on the surface density oscillation in the intrinsic states.
The intrinsic states of $^{12}$C and $^{16}$O obtained by the AMD+VAP method show 
triangle and tetrahedral shapes, respectively, because of the $\alpha$ correlations.
The formation of $\alpha$ clusters in these states was confirmed in the AMD framework,
in which existence of any clusters are not {\it a priori} assumed.
The intrinsic deformations are regarded as 
spontaneous symmetry breaking of rotational invariance. It was shown that the 
oscillating surface density in the triangle and tetrahedral shapes is associated with 
that in DW states caused by the instability of Fermi surface with 
respect to a kind of $1p$-$1h$ correlations.

To discuss the symmetry breaking between uniform density states and the oscillating density states, 
a schematic model of a few clusters on a Fermi gas core in a one-dimensional finite box 
was introduced.  In the model analysis, we conjecture structure transitions from 
a Fermi gas state to a DW-like state via a BCS-like state, and to a BEC-like state 
depending on the cluster size relative to the box size. 

In both analyses of the BB-cluster model and the schematic 1D-cluster model,
Pauli blocking effects are found to play an important role in the DW-like state.
The breaking of the translational invariance in the DW-like state 
originates in the Pauli blocking effect between clusters, which acts as an effective 
inter-cluster repulsion and restricts cluster motion.

In the present analysis with the schematic 1D-cluster model, 
we do not perform energy variation nor calculate energy eigenstates.
Moreover, the mechanism of cluster formation and the dynamical change of 
the cluster structure or the cluster size are beyond our scope in the present paper. 
%Based on simple anzats, we give a model wave function with fixed parameters $b$ and $\tilde k_F$ for 
%the cluster size relative to the box size and the lowest allowed momentum, 
%and analyze Pauli blocking effect between a cluster and the 
%core and that between two clusters in the given wave function. 
The extension of the present model by taking into account
dynamical change of cluster structure with the use of effective nuclear interactions is an 
important issue to be solved in future study. 
Also the cluster and core formations as well as diffuseness of the core Fermi surface should be
studied in more realistic models. 
Furthermore, the assumption that the interaction between clusters is weak and 
the Pauli blocking effect gives the major contribution to 
the inter-cluster motion may be too simple.
To clarify which state of BCS-like, DW-like, or EXc-like ones realizes 
it is essential to explicitly solve the problem of inter-cluster motion 
by taking into account nuclear forces or inter-cluster interactions.

It would be interesting to associate the present picture for clusters in the 1D finite box 
with phase transitions in infinite matter. 
In the extension of the present model to infinite matter problem,
one should take care of the differences between finite systems and infinite systems as follows.
Firstly, momentum $k$ values in a finite box is discrete because of boundary condition, 
while they are continuum values in infinite matter. 
In the description with discrete momentum, 
long range correlations beyond the box size $L$ is not taken into account.
What we call the BCS-type correlation in the present 1D-cluster model is 
the correlation in the range of the box size at most.
In second, the total momentum of c.m.m. should be projected to zero in finite system, 
while it is not necessarily zero in infinite systems. 
In spite of those differences, the 1D-cluster model 
may give a hint to understand an origin of DW in infinite matter.

%\begin{figure}[th]\label{fig:kf3}
%\centerline{\epsfxsize=3.7 cm\epsffile{k3-b0.75-k-p.eps}\epsfxsize=4.5 %cm\epsffile{k3-b0.75-rho1-p.eps}\epsfxsize=4.5 cm\epsffile{k3-b0.75-rho2-p.eps}}
%\centerline{\epsfxsize=3.7 cm\epsffile{k3-b0.75-dif-k-p.eps}\epsfxsize=4.5 %cm\epsffile{k3-b0.75-dif-rho1-p.eps}\epsfxsize=4.5 cm\epsffile{k3-b0.75-dif-rho2-p.eps}}
%\caption{Same as Fig.~\ref{fig:kf1} but for $\tilde k_F=3$ without and with the diffuseness of the %core Fermi surface. The clusters size $\tilde b$ is $\tilde b=0.75$.
%}
%\end{figure}

\section*{Acknowledgments}
The authors thank to nuclear theory group members of department of physics of Kyoto University 
for 
valuable discussions. 
Discussions during the YIPQS long-term workshop "DCEN2011" held at YITP are 
helpful to advance this work.
The computational calculations of this work were performed by using the
supercomputers at YITP.
This work was supported by Grant-in-Aid for Scientific Research from Japan Society for the Promotion of Science (JSPS) 
Grant Number [No.23340067, 24740184].
It was also supported by 
the Grant-in-Aid for the Global COE Program ``The Next Generation of Physics, 
Spun from Universality and Emergence'' from the Ministry of Education, Culture, Sports, Science and Technology (MEXT) of Japan. 

\appendix
\section{Triangle and tetrahedral deformations in the $3\alpha$- and $4\alpha$-cluster structures}\label{app:BB-MF}
We here describe triangle and tetrahedral deformations and particle-hole representations of $3\alpha$- and $4\alpha$-cluster wave functions.
As explained in Ref.~\cite{KanadaEn'yo:2011qf}, the Brink-Bloch (BB) $3\alpha$-cluster wave 
functions~\cite{brink66} with a regular triangle 
configuration in the small inter-cluster distance case can be rewritten as, 
\begin{equation}
\Phi^{BB}_{3\alpha}(\epsilon) \approx \prod_{\tau\sigma}\left\{ \phi_{00}{\cal X}_{\tau\sigma}
(\phi_{1-1}+\epsilon\phi_{2+2}){\cal X}_{\tau\sigma}
(\phi_{1+1}-\epsilon\phi_{2-2}){\cal X}_{\tau\sigma}\right\}.
\end{equation}
Here $\phi_{lm}$ is the single-particle orbit in the harmonic oscillator potential,
${\cal X}_{\tau\sigma}$ is the spin-isospin wave function with 
$\tau=\{p,n\}$ and $\sigma=\{\uparrow,\downarrow\}$.
$\epsilon$ is the order parameter for the axial symmetry breaking and it corresponds to $u/v$ in the BCS theory. Here $\epsilon$ is considered to be small to omit $\epsilon^2$ and higher terms and 
taken to be a real value.
In principle, $\epsilon$ can be a complex value but the phase is arbitrary and can be changed by 
the rotation around the $z$ axis (transformation of the polar 
coordinate $\varphi$).
The density of the $\Phi^{BB}_{3\alpha}(\epsilon)$ state is 
\begin{equation}
\rho({\bf r})=\frac{4}{\pi^{3/2}b^3}e^{-\frac{r^2}{b^2}}\left\{ 1+2\frac{r^2}{b^2}\sin^2(\theta)
+\epsilon 2\sqrt{2}\frac{r^3}{b^3}\sin^3(\theta)(e^{-3i\varphi}-e^{3i\varphi})+{\cal O}(\epsilon^2) \right\}, 
\end{equation}
and its multipole decomposition at a certain radius $r=R_0$ is
\begin{equation} \label{eq:3alpha-density}
\rho(R_0, \theta, \varphi) = \frac{8}{\pi^{1/2}b^3}e^{-\frac{R_0^2}{b^2}}
\left\{(1+\frac{4}{3}\frac{R_0^2}{b^2})Y_0^0(\theta,\varphi) 
-\frac{4}{3\sqrt{5}}\frac{R_0^2}{b^2}Y_2^0(\theta,\varphi)
+\epsilon \frac{8}{\sqrt{35}}\frac{R_0^3}{b^3}
\left(\frac{Y_3^{-3}(\theta,\varphi)}{\sqrt{2}}-\frac{Y_3^{+3}(\theta,\varphi)}{\sqrt{2}}\right)
+{\cal O}(\epsilon^2)\right\}.
\end{equation}

In a similar way, 
the BB $4\alpha$-cluster wave functions with a regular tetrahedral
configuration can be rewritten as
\begin{equation}
\Phi^{BB}_{4\alpha}(\epsilon) \approx \prod_{\tau\sigma}\left\{ \phi_{00}{\cal X}_{\tau\sigma}
\phi_{10}{\cal X}_{\tau\sigma}
(\phi_{1-1}+\epsilon\phi_{2+1}){\cal X}_{\tau\sigma}
(\phi_{1+1}+\epsilon\phi_{2-1}){\cal X}_{\tau\sigma}\right\}.
\end{equation}
Here it is assumed that the inter-cluster distance in the tetrahedral configuration is small, i.e., 
$\epsilon$ is small. 
The density is 
\begin{equation}
\rho({\bf r})=\frac{4}{\pi^{3/2}b^3}e^{-\frac{r^2}{b^2}}\left\{1+2\frac{r^2}{b^2}\sin^2(\theta)
+\epsilon \sqrt{2}\frac{r^3}{b^3}\sin^2(\theta)(e^{2i\varphi}+e^{-2i\varphi})+{\cal O}(\epsilon^2) \right\}, 
\end{equation}
and its multipole decomposition at a radius $r=R_0$ is
\begin{equation}\label{eq:4alpha-density}
\rho(R_0, \theta, \varphi) = \frac{8}{\pi^{1/2}b^3}e^{-\frac{R_0^2}{b^2}}
\left\{(1+2\frac{R_0^2}{b^2})Y_0^0(\theta,\varphi) 
+\epsilon \sqrt{\frac{32}{105}}\frac{R_0^3}{b^3}(\frac{Y_3^{-2}(\theta,\varphi)}{\sqrt{2}}+
\frac{Y_3^{+2}(\theta,\varphi)}{\sqrt{2}})
\right\}.
\end{equation}
$Y_3^{-2}(\theta,\varphi)/\sqrt{2}+Y_3^{+2}(\theta,\varphi)/\sqrt{2}$ in the $\epsilon$ term
can be also transformed to  
\begin{equation}\label{eq:4alpha-density2}
\frac{Y_3^{-2}(R_\Omega\theta',R_\Omega\varphi')}{\sqrt{2}}+\frac{Y_3^{+2}(\theta',\varphi')}{\sqrt{2}}
= \frac{\sqrt{5}}{3}Y_3^0(\theta,\varphi)+\frac{\sqrt{2}}{3}Y_3^{-3}(\theta,\varphi)
-\frac{\sqrt{2}}{3}Y_3^{+3}(\theta,\varphi), 
\end{equation}
by a $\Omega$ rotation $R_\Omega$.%,where $(\theta',\varphi')=R(\Omega)(\theta,\varphi)$.

By using the creation and annihilation operators, $a^\dagger_{lm,\chi}$ and
$a_{lm,\chi}$ for the $\phi_{lm}{\cal X}_{\tau\sigma}$ state with $\chi=\tau\sigma$, 
the $\Phi^{BB}_{3\alpha}(\epsilon)$ and $\Phi^{BB}_{4\alpha}(\epsilon)$ states are expressed as
\begin{align}
| \Phi^{BB}_{3\alpha}(\epsilon)\rangle &= \prod_{\chi} 
a^\dagger_{00,\chi} (a_{1-1,\chi}+\epsilon a^\dagger_{2+2,\chi})
(a_{1+1,\chi}-\epsilon a^\dagger_{2-2,\chi}) |-\rangle, \\
| \Phi^{BB}_{4\alpha}(\epsilon)\rangle &= \prod_{\chi} 
a^\dagger_{00,\chi}a^\dagger_{10,\chi}  (a_{1-1,\chi}+\epsilon a^\dagger_{2+1,\chi})
(a_{1+1,\chi}+\epsilon a^\dagger_{2-1,\chi}) |-\rangle.
\end{align}
In the particle and hole representation, they are rewritten as 
\begin{align}\label{eq:3alpha-ph}
| \Phi^{BB}_{3\alpha}(\epsilon)\rangle &= \prod_{\chi} 
(1+\epsilon a^\dagger_{2+2,\chi}b^\dagger_{1+1,\chi})
(1-\epsilon a^\dagger_{2-2,\chi}b^\dagger_{1-1,\chi}) |0\rangle_F,\\
|0\rangle_F&\equiv
 \prod_{\chi} 
\left( a^\dagger_{00,\chi} a_{1-1,\chi} a_{1+1,\chi}\right )  |-\rangle,
\end{align}
and 
\begin{align}\label{eq:4alpha-ph}
| \Phi^{BB}_{4\alpha}(\epsilon)\rangle &= \prod_{\chi} 
(1+\epsilon a^\dagger_{2+1,\chi}b^\dagger_{1+1,\chi})
(1+\epsilon a^\dagger_{2-1,\chi}b^\dagger_{1-1,\chi}) |0\rangle_F,\\
|0\rangle_F &\equiv
\prod_{\chi} 
\left( a^\dagger_{00,\chi} a^\dagger_{10,\chi}  a_{1-1,\chi} a_{1+1,\chi}\right )  |-\rangle.
\end{align}
Here the hole operator $b^\dagger_{lm,\chi}$ is defined to be
$b^\dagger_{lm,\chi}=a_{l-m,\chi}$. 
Equations~(\ref{eq:3alpha-ph}) and (\ref{eq:4alpha-ph}) indicate that 
the $3\alpha$- and $4\alpha$-cluster wave functions contains 
the DW-type $1p$-$1h$ correlations carrying finite angular momenta.

\section{Weak coupling regime of one-dimensional cluster model} \label{sec:weak-coupling1}
\label{app:weak-coupling}

We consider the weak-coupling case of the 1D-cluster model described in \ref{sec:1D-cluster}.
Hereafter, we take $r_0=1$ and consider the case that 
all the parameters equal to the dimensionless parameters, for instance, 
$b=\tilde b$, $k=\tilde k$, and $s=\tilde s$. 
We define the creation and annihilation operators $a^\dagger_{k\chi}$ and $a^\dagger_{-k\chi}$ 
for the single-particle states $\phi_k{\cal X}_\chi$ and $\phi_k{\cal X}_\chi$, where 
$\phi_k$ and $\phi_{-k}$ are momentum $k$ and $-k$ plane waves written as
$\phi_k(x)=e^{+ik x}/{\sqrt{2L}}$ and  $\phi_{-k}(x)=e^{-ikx}/{\sqrt{2L}}$ in the 
coordinate representation.
The core state $|C\rangle$ where all $k < k_F$ states are occupied by core nucleons
is written as 
\begin{equation}
|C\rangle = \prod_\chi\prod_{|k|<k_F}  a^\dagger_{k_F\chi} a^\dagger_{-k_F\chi} |-\rangle,
\end{equation}
where $|-\rangle$ is the vacuum.

In the case of weak coupling, the coefficient $f(k)$  of the 
momentum $k$ and $-k$ plane waves in Eq.~(\ref{eq:psi_1}) is dominant for $k=k_F$ and 
it rapidly decreases as $k$ increases.
In the limit, $|f(k_F)| \gg |f_(k_F+1)| \gg  |f_(k_F+2)| \cdots $, 
the leading term of the $2\alpha$-cluster state in the 1D-cluster model 
is the Fermi gas state with the surface momentum $k_F$,
which is the uncorrelated state. For the correlated system, 
the deviation from the Fermi gas state is important. We take into account 
the next leading term,
$k=k_F+1$ components, and ignore the higher momentum components for $k \ge k_F+2$.
We set $f(k_F)=1$ and $f(k_F+1)=\epsilon$ with $\epsilon << 1$ and take 
the leading term of $\epsilon$. Then the state consisting of two $\alpha$ clusters around $x=0$ and $x=s$ 
written by
Eqs.~(\ref{eq:psi_1}) and (\ref{eq:psi_2s}) on the Fermi gas core is expressed as
\begin{align}
|\psi_{\alpha_1}\psi^s_{\alpha_2} C\rangle &\propto  \prod_\chi 
\left\{ (c^*_0-c_0)  a^\dagger_{k_F\chi} a^\dagger_{-k_F\chi} \right. \notag\\
&\quad+ \epsilon (c_0-c_1) a^\dagger_{k'_F\chi}  a^\dagger_{k_F\chi} 
+ \epsilon (c^*_0-c^*_1) a^\dagger_{-k'_F\chi}  a^\dagger_{-k_F\chi} \notag\\
&\quad+ \epsilon (c^*_0-c_1) a^\dagger_{k'_F\chi}  a^\dagger_{-k_F\chi} 
+ \epsilon(c_0-c^*_1) a^\dagger_{-k'_F\chi}  a^\dagger_{k_F\chi} \notag\\
&\quad+ \left. \epsilon^2 (c^*_1-c_1) a^\dagger_{k'_F\chi}  a^\dagger_{-k'_F\chi} 
+\cdots  \right\} |C\rangle, 
\end{align}
where $k'_F=k_F+1$, $c_0(s)=e^{-ik_Fs}$, and $c_1(s)=e^{-ik'_Fs}$.
The first term is the Fermi gas state with the Fermi surface at $k_F$. 
Its coefficient $c^*_0-c_0=2i \sin(k_Fs)$ vanishes 
for $s=\pi m/k_F$ ($m$ is an integer) because of Pauli principle.
In other words, the area around $s=\pi 2m/2k_F$ ($m=0,\cdots,2 k_F$) is forbidden 
for the $\alpha_2$ cluster center position,
while the area around $s_j=\pi (2j-1)/2k_F$ ($j=1,\cdots,2 k_F$) is allowed. 

In the particle-hole representation with respect to the Fermi surface at $k_F$, 
the second and third terms correspond to 
the $1p-1h$ states with the DW-type correlation with the wave number $k'_F+k_F=2k_F+1$, 
and the forth and fifth terms are those with the Exc-type correlation 
with the wave number
$k'_F-k_F=1$, and the last term is the $2p$-$2h$ state. We rewrite the coefficients in 
$|\psi_{\alpha_1}\psi^s_{\alpha_2} C\rangle$ as
\begin{align}
|\psi_{\alpha_1}\psi^s_{\alpha_2} C\rangle &= n^2_0 \prod_\chi 
C_0(s) \left\{  a^\dagger_{k_F\chi} a^\dagger_{-k_F\chi} \right. \notag\\
&\quad+ \epsilon C_{DW}(s) a^\dagger_{k'_F\chi}  a^\dagger_{k_F\chi} 
+ \epsilon C^*_{DW}(s) a^\dagger_{-k'_F\chi}  a^\dagger_{-k_F\chi}  \notag\\
&\quad+ \epsilon C_{Exc}(s) a^\dagger_{k'_F\chi}  a^\dagger_{-k_F\chi} 
+  \epsilon C^*_{Exc}(s) a^\dagger_{-k'_F\chi}  a^\dagger_{k_F\chi}  \notag\\
&\quad+ \left. \epsilon^2 C_2(s)  a^\dagger_{k'_F\chi}  a^\dagger_{-k'_F\chi} 
+\cdots  \right\} |C\rangle,\\
&\quad C_0(s)\equiv c^*_0-c_0=-2i \sin(k_Fs),\\
&\quad C_{DW}(s) \equiv  \frac{c_0-c_1}{c^*_0-c_0} = \frac{e^{-is}-1}{2},\\
&\quad C_{Exc}(s) \equiv  \frac{c^*_0-c_1}{c^*_0-c_0} = \frac{e^{-is}+1}{2}, \\
&\quad C_2(s) \equiv \frac{c^*_1-c_1}{c^*_0-c_0} = \cos(is).
\end{align}
By using the hole operator $b^\dagger_{k\chi}= a_{-k\chi}$ we obtain 
\begin{align}
|\psi_{\alpha_1}\psi^s_{\alpha_2} C\rangle &= n^2_0 \prod_\chi 
C_0(s) \left\{  1 \right. \notag\\
&\quad- \epsilon C_{DW}(s) a^\dagger_{k'_F\chi}  b^\dagger_{k_F\chi} 
+ \epsilon C^*_{DW}(s) a^\dagger_{-k'_F\chi}  b^\dagger_{-k_F\chi} \notag\\
&\quad+ \epsilon C_{Exc}(s) a^\dagger_{k'_F\chi}  b^\dagger_{-k_F\chi} 
-  \epsilon C^*_{Exc}(s) a^\dagger_{-k'_F\chi}  b^\dagger_{k_F\chi} \notag\\
&\quad+ \left. \epsilon^2 C_2(s)  a^\dagger_{k'_F\chi}  a^\dagger_{-k'_F\chi}b^\dagger_{-k_F\chi}  b^\dagger_{k_F\chi} 
+\cdots  \right\} |HF \rangle,
\end{align}
where the Hartree-Fock vacuum $|HF \rangle\equiv \prod_\chi  a^\dagger_{k_F\chi} a^\dagger_{-k_F\chi} |C\rangle$.
When $s$ is the allowed $s_j$ close to $L/2=\pi$, i.e., the $\alpha_2$ center is located around the middle of the box, $e^{-is}$ approaches $-1$, and therefore, 
the DW correlation becomes dominant and the Exc correlation is minor, while 
they are opposite for $s_j$ close to $0$.
In both cases, the correlation disappears in the $\epsilon\rightarrow 0$ limit, 
and the system goes to the Fermi gas limit.

\section{Transition between DW/Exc-like and BCS-like states}\label{sec:weak-coupling2}
We extend the model in the weak coupling limit described in the previous section
by superposing all allowed $s_j$ states for the $\alpha_2$ wave function, 
\begin{align}
\psi_{\alpha_2}&=\sum_j F(s_j) \psi^{s_j}_{\alpha_2},\\
|\Psi\rangle&= \sum_jF(s_j)| \psi_{\alpha_1}\psi^{s_j}_{\alpha_2} C\rangle.
\end{align}
In principle, one should determine the coefficients $F(s_j)$ following the energy variation
to properly take into account dynamics of relative motion between clusters, $\alpha_1$ and $\alpha_2$.
However, we here do not perform energy optimization but  
adopt a simple ansatz for $F(s_j)$ as follows.

As described before, if the $s_j\sim L/2$ ($s_j\sim 0$) components are main, 
the DW (Exc) correlation is dominant. In more general, in case that the effective 
interaction between $\alpha$ clusters is repulsive, the amplitudes may be relatively 
smaller in the region around $s_j\sim 0$ (and also around $s_j\sim L$ due to the periodic boundary) 
than that around $s_j= L/2=\pi$ resulting in the DW dominance. 
On the other hand, if the interaction between $\alpha$ clusters is attractive, 
the amplitudes concentrate on the region around $s_j\sim 0$, which may cause the Exc dominance.
In both cases, spatial correlation between $\alpha_1$ and $\alpha_2$ remains and the system shows
density oscillation and the symmetry breaking of translational invariance.

Next we construct the state where there is no spatial 
correlation between $\alpha_1$ and $\alpha_2$ (no inter-cluster correlation)
and the translational invariance is completely 
restored by the superposition and the $K_G=0$ (total momentum of c.m.m.)
projection of the $|\psi_{\alpha_1}\psi^{s_j}_{\alpha_2} C\rangle$. 
The state with no spacial correlation between $\alpha_1$ and $\alpha_2$  can be constructed by 
the projection to the relative momentum $k_{\alpha_1}-k_{\alpha_2}=0$ state. Here $k_{\alpha_{1,2}}$
stands for the momentum of c.m.m. of $\alpha_{1,2}$. 
In the weak coupling regime, this is performed by superposing all allowed $s_j$ states 
with an equal weight as 
\begin{equation}
\psi_{\alpha_2}=\sum_j \frac{1}{(-2i(-1)^j)^{n}} \psi^{s_j}_{\alpha_2}.
\end{equation}
Taking averages of $C_{DW}(j)$, $C_{Exc}(j)$ and $C_2(j)$ with respect to $j$, we obtain 
%\begin{eqnarray}
%|\psi_{\alpha_1}\psi_{\alpha_2} C\rangle &\propto & \prod_\chi 
%\left\{ a^\dagger_{k_F\chi} a^\dagger_{-k_F\chi} \right. \\
%&&-\frac{\epsilon}{2} (a^\dagger_{k'_F\chi}  a^\dagger_{k_F\chi} 
%+ a^\dagger_{-k'_F\chi}  a^\dagger_{-k_F\chi} ) \\
%&&\left. +\frac{\epsilon}{2} ( a^\dagger_{k'_F\chi}  a^\dagger_{-k_F\chi} 
%+ a^\dagger_{-k'_F\chi}  a^\dagger_{k_F\chi}) +\cdots \right\} |C\rangle
%\end{eqnarray}
%By using the hole operator $b^\dagger_{k\chi}= a_{-k\chi}$ we get 
\begin{align}
|\psi_{\alpha_1}\psi_{\alpha_2} C\rangle &\propto  \prod_\chi 
\left\{ 1 - \frac{\epsilon}{2} (a^\dagger_{k'_F\chi}  b^\dagger_{k_F\chi} 
- a^\dagger_{-k'_F\chi}  b^\dagger_{-k_F\chi} ) \right. \notag\\
&\quad\left. - \frac{\epsilon}{2} (a^\dagger_{k'_F\chi}  b^\dagger_{-k_F\chi} 
- a^\dagger_{-k'_F\chi}  b^\dagger_{k_F\chi}) +\cdots \right\} |HF \rangle,
\end{align}
%where the Hartree-Fock vacuum $|HF \rangle\equiv \prod_\chi  a^\dagger_{k_F\chi} a^\dagger_{-k_F\chi} |C\rangle$.
This corresponds to the intrinsic state for no correlation between clusters. In the ground state of 
the finite system, the total momentum $K_G=k_{\alpha_1}+k_{\alpha_2}$ should be 
zero.
With the projection onto $K_G=0$, 
we finally obtain the symmetry restored state, 
\begin{equation}\label{eq:BCS}
P^{K_G=0}|\psi_{\alpha_1}\psi_{\alpha_2} C\rangle
\propto \left\{ 1+\frac{\epsilon^2}{4}\sum_{\chi_\alpha\chi_\beta} 
(a^\dagger_{k'_F\chi_\alpha}a^\dagger_{-k'_F\chi_\beta} -
a^\dagger_{k'_F\chi_\beta}a^\dagger_{-k'_F\chi_\alpha} )
(b^\dagger_{k_F\chi_\alpha}b^\dagger_{-k_F\chi_\beta} -
b^\dagger_{k_F\chi_\beta}b^\dagger_{-k_F\chi_\alpha} ) +\cdots \right\} |HF \rangle .
\end{equation}
It is found that the state contains $2p$-$2h$ configurations and it is consistent with 
a BCS-like state, where two particles form a 
$(k'_F,-k'_F)$ $\chi_\alpha\chi_\beta$ pair and two holes do a $(k_F,-k_F)$ $\chi_\alpha\chi_\beta$ pair. 
In the $2p$-$2h$ state, all kinds of $\chi$ pairing is coherently mixed so as to keep 
the spin-isospin symmetry originating in the symmetry of $\alpha$ clusters.

Note that, in the inter-cluster correlated case of the DW dominance state, the $K_G=0$ projected state
contains 
$a^\dagger_{k'_F\chi_\alpha}a^\dagger_{-k'_F\chi_\beta}
b^\dagger_{k_F\chi_\alpha}b^\dagger_{-k_F\chi_\beta}+
a^\dagger_{k'_F\chi_\beta}a^\dagger_{-k'_F\chi_\alpha}
b^\dagger_{k_F\chi_\beta}b^\dagger_{-k_F\chi_\alpha}$ in the $\epsilon^2$ term,
which means that $1p$-$1h$ of $\chi_\alpha$ always carries the finite momentum $\pm(k'_F+k_F)$
as a result of spatial correlation between clusters.


\section*{References}
\begin{thebibliography}{9}
\bibitem{wheeler37}
J. A. Wheeler, Phys. Rev. {\bf 52}, 1083 (1937); {\it ibid.} {\bf 52}, 1107 (1937).

\bibitem{dennison54}
D. M. Dennison, Phys. Rev. {\bf 96}, 378 (1954). 

\bibitem{brink70}
D. M. Brink, H. Friedrich, A. Weiguny and C. W. Wong, Phys. Lett. {\bf B33}, 143 (1970). 

\bibitem{ikeda72}
K. Ikeda {\it et al.}, Prog. Theor. Phys. Suppl. {\bf 52}, 1 (1972).



\bibitem{OCM}
H. Horiuchi, Prog. Theor. Phys. {\bf 51}, 1266 (1974); {\bf 53}, 447 (1975).
 
\bibitem{smirnov74}
Yu. F. Smirnov, I. T. Obukhovsky. Yu. M. Tchuvil'sky and V. G. Neudatchin, Nucl. Phys. A 235, 289 (1974).

\bibitem{GCM}
E. Uegaki, S. Okabe, Y. Abe and H. Tanaka, Prog. Theor. Phys. {\bf 57},
1262 (1977).
E. Uegaki, Y. Abe, S. Okabe and H. Tanaka, Prog. Theor. Phys. {\bf 59},
 1031 (1978); {\bf 62}, 1621 (1979).

\bibitem{RGM}
Y. Fukushima and M. Kamimura,
{\it Proc. Int. Conf. on Nuclear Structure, Tokyo, 1977,
edited by T. Marumori}[J. Phys. Soc. Jpn. {\bf 44}, 225 (1978);
M. Kamimura, Nucl. Phys. {\bf A351}, 456 (1981).

\bibitem{suzuki76} %OCM for 16O
Y. Suzuki, Prog. Theor. Phys. {\bf 55}, 1751 (1976). 


\bibitem{fujiwara80}
Y. Fujiwara {\em et al.}, Prog. Theor. Phys. Suppl.{\bf 68}, 29 (1980).

\bibitem{bauhoff84}
W. Bauhoff, H. Schultheis, R. Schultheis
Phys. Rev. {\bf C 29}, 1046 (1984).


\bibitem{abe71}
Y. Abe, J. Hiura and H. Tanaka, Prog. Theor. Phys. {\bf 46}, 352 (1971); 
{\it ibid.} {\bf 49}, 800 (1972).


\bibitem{takigawa71}
N. Takigawa and A. Arima, Nucl. Phys. A {\bf 168}, 593 (1971).
 
\bibitem{Tohsaki01}
A. Tohsaki, H. Horiuchi, P. Schuck, and G. R\"opke, 
Phys. Rev. Lett. {\bf 87}, 192501 (2001). 

\bibitem{Ropke98}
G. R\"opke, A. Schnell, P. Schuck, and P. Nozieres, Phys. Rev. Lett. {\bf 80},
3177 (1998).


\bibitem{overhauser60}
A. W. Overhauser, Phys. Rev. Lett. {\bf 4}, 415 (1960).
\bibitem{brink73}  D. M. Brink and J. J. Castro, Nucl. Phys. {\bf A216}, 109 (1973).
\bibitem{llano79} M. de Llano, Nucl. Phys. A {\bf 317}, 183 (1979).
\bibitem{ui81} H. Ui and Y. Kawazore, Z. Phys. A {\bf 301}, 125 (1981). 
\bibitem{tamagaki76}
R. Tamagaki and T. Takatsuka, Prog. Theor. Phys. {\bf 56},1340 (1976).
\bibitem{takatsuka78}
T. Takatsuka, K. Tamiya, T. Tatsumi and R. Tamagaki, Prog. Theor. Phys. {\bf 59}, 1933 (1978). 
\bibitem{migdal78}
A. B. Migdal, Rev. Mod. Phys. {\bf 50}, 107 (1978).

\bibitem{Dautry:1979bk}
  F.~Dautry and E.~M.~Nyman,
  %``Pion Condensation And The Sigma Model In Liquid Neutron Matter,''
  Nucl.\ Phys.\  A {\bf 319}, 323 (1979).
  %%CITATION = NUPHA,A319,323;%%

\bibitem{Deryagin:1992rw}
  D.~V.~Deryagin, D.~Y.~Grigoriev and V.~A.~Rubakov,
  %``Standing wave ground state in high density, zero temperature QCD at large
  %N(c),''
  Int.\ J.\ Mod.\ Phys.\  A {\bf 7}, 659 (1992).
  %%CITATION = IMPAE,A7,659;%%

\bibitem{Shuster:1999tn}
  E.~Shuster and D.~T.~Son,
  %``On finite density QCD at large N(c),''
  Nucl.\ Phys.\  B {\bf 573}, 434 (2000).
%  [arXiv:hep-ph/9905448].
  %%CITATION = NUPHA,B573,434;%%

\bibitem{Park:1999bz}
  B.~Y.~Park, M.~Rho, A.~Wirzba and I.~Zahed,
  %``Dense QCD: Overhauser or BCS pairing?,''
  Phys.\ Rev.\  D {\bf 62}, 034015 (2000).
%  [arXiv:hep-ph/9910347].
  %%CITATION = PHRVA,D62,034015;%%

\bibitem{Alford:2000ze}
  M.~G.~Alford, J.~A.~Bowers and K.~Rajagopal,
  %``Crystalline color superconductivity,''
  Phys.\ Rev.\  D {\bf 63}, 074016 (2001).
  %[arXiv:hep-ph/0008208].
  %%CITATION = PHRVA,D63,074016;%%


\bibitem{Nakano:2004cd}
  E.~Nakano and T.~Tatsumi,
  %``Chiral symmetry and density wave in quark matter,''
  Phys.\ Rev.\  D {\bf 71}, 114006 (2005).
  %[arXiv:hep-ph/0411350].
  %%CITATION = PHRVA,D71,114006;%%
\bibitem{Giannakis:2004pf}
  I.~Giannakis and H.~C.~Ren,
  %``Chromomagnetic instability and the LOFF state in a two flavor color
  %superconductor,''
  Phys.\ Lett.\  B {\bf 611}, 137 (2005).
%  [arXiv:hep-ph/0412015].
  %%CITATION = PHLTA,B611,137;%%

\bibitem{Fukushima:2006su}
  K.~Fukushima,
  %``Characterizing the Larkin-Ovchinnikov-Fulde-Ferrel phase induced by the
  %chromomagnetic instability,''
  Phys.\ Rev.\  D {\bf 73}, 094016 (2006).
%  [arXiv:hep-ph/0603216].
  %%CITATION = PHRVA,D73,094016;%%

\bibitem{Nickel:2009ke}
  D.~Nickel,
  %``How many phases meet at the chiral critical point?,''
  Phys.\ Rev.\ Lett.\  {\bf 103}, 072301 (2009);
%  [arXiv:0902.1778 [hep-ph]];
  %%CITATION = PRLTA,103,072301;%%
  %``Inhomogeneous phases in the Nambu-Jona-Lasino and quark-meson model,''
  Phys.\ Rev.\  D {\bf 80}, 074025 (2009).
%  [arXiv:0906.5295 [hep-ph]].
  %%CITATION = PHRVA,D80,074025;%%
 
\bibitem{Kojo:2009ha}
  T.~Kojo, Y.~Hidaka, L.~McLerran and R.~D.~Pisarski,
  %``Quarkyonic Chiral Spirals,''
  Nucl.\ Phys.\  A {\bf 843}, 37 (2010).
  %[arXiv:0912.3800 [hep-ph]].
  %%CITATION = NUPHA,A843,37;%%

%\cite{Carignano:2010ac}
\bibitem{Carignano:2010ac}
  S.~Carignano, D.~Nickel and M.~Buballa,
  %``Influence of vector interaction and Polyakov loop dynamics on inhomogeneous
  %chiral symmetry breaking phases,''
  Phys.\ Rev.\  D {\bf 82}, 054009 (2010).
%  [arXiv:1007.1397 [hep-ph]].
  %%CITATION = PHRVA,D82,054009;%%

%\cite{Schon:2000qy}
%\bibitem{Schon:2000qy}
%  V.~Schon and M.~Thies,
%  %``2-D model field theories at finite temperature and density,''
%  arXiv:hep-th/0008175.
  %%CITATION = HEP-TH/0008175;%%


  

\bibitem{Fukushima:2010bq}
  K.~Fukushima, T.~Hatsuda,
  %``The phase diagram of dense QCD,''
  Rept.\ Prog.\ Phys.\  {\bf 74}, 014001 (2011).
%\bibitem{Gete:2000yd}
%  E.~Gete {\it et al.},
%  %``Beta Delayed Particle Decay Of 9c And The A=9, T=1/2 Nuclear System:
%  %Experiment, Data, And Phenomenological Analysis,''
%  Phys.\ Rev.\  C {\bf 61}, 064310 (2000).
%  %%CITATION = PHRVA,C61,064310;%%


\bibitem{CDW}
 G. Gruner, Rev. Mod. Phys. 60, 1129 (1988). 
%The dynamics of charge-density waves,
 \bibitem{SDW}
G. Gruner, Rev. Mod. Phys. 66, 1 (1994).
%The dynamics of spin-density waves,

\bibitem{KanadaEn'yo:2011qf} 
  Y.~Kanada-En'yo and Y.~Hidaka,
  %``Alpha-cluster structure and density wave in oblate nuclei,''
  Phys.\ Rev.\ C {\bf 84}, 014313 (2011).
%  [arXiv:1104.4140 [nucl-th]].
  %%CITATION = ARXIV:1104.4140;%%

\bibitem{eichler70}
J. Eichler, A. Faessler
Nucl. Phys., A157,166 (1970).

\bibitem{onishi71}
N. Onishi, R.K. Sheline
Nucl. Phys. A165, 180 (1971).

\bibitem{takami95}
S. Takami, K. Yabamna, K. Ikeda, Prog. Theor. Phys. {\bf 96}, 407 (1996).

\bibitem{Robson:1979zz} 
  D.~Robson,
  %``Evidence for the Tetrahedral Nature of O-16,''
  Phys.\ Rev.\ Lett.\  {\bf 42}, 876 (1979).
  %%CITATION = PRLTA,42,876;%%

\bibitem{elliott85}
J.P.Elliott, J.A.Evans, E.E.Maqueda
Nucl.Phys. A437, 208 (1985).





  



\bibitem{sheline60}
R. K. Sheline and K. Wildermuth, Nucl. Phys. 21, 196 (1960).

\bibitem{horiuchi68} %parity-doublet
H. Horiuchi and K. Ikeda, Prog. Theor. Phys. 40 (1968), 277[PTP]. 





\bibitem{Funaki:2008gb} 
  Y.~Funaki, T.~Yamada, H.~Horiuchi, G.~Ropke, P.~Schuck and A.~Tohsaki,
  %``Alpha-particle condensation in O-16 via a full four-body OCM calculation,''
  Phys.\ Rev.\ Lett.\  {\bf 101}, 082502 (2008).
%  [arXiv:0802.3246 [nucl-th]].
  %%CITATION = ARXIV:0802.3246;%%
  
\bibitem{Funaki:2010px} 
  Y.~Funaki, T.~Yamada, A.~Tohsaki, H.~Horiuchi, G.~Ropke and P.~Schuck,
  %``Microscopic study of 4-alpha-particle condensation with proper treatment of resonances,''
  Phys.\ Rev.\ C {\bf 82}, 024312 (2010).
%  [arXiv:1004.2310 [nucl-th]].
  %%CITATION = ARXIV:1004.2310;%%  


\bibitem{ENYOabc}
  Y.~Kanada-En'yo and H.~Horiuchi,
  %``Clustering in yrast states of Ne-20 studied with antisymmetrized molecular
  %dynamics,''
  Prog.\ Theor.\ Phys.\  {\bf 93}, 115 (1995);
  %%CITATION = PTPKA,93,115;%%
%\bibitem{KanadaEnyo:1995tb}
  Y.~Kanada-En'yo, H.~Horiuchi and A.~Ono,
  %``Structure of Li and Be isotopes studied with antisymmetrized molecular
  %dynamics,''
  Phys.\ Rev.\  C {\bf 52}, 628 (1995);
  %%CITATION = PHRVA,C52,628;%%
%\bibitem{KanadaEnyo:1995ir}
  Y.~Kanada-En'yo and H.~Horiuchi,
%  %``Neutron-Rich B Isotopes Studied With Antisymmetrized Molecular Dynamics,''
  Phys.\ Rev.\  C {\bf 52}, 647 (1995).
%  %%CITATION = PHRVA,C52,647;%%
  
\bibitem{ENYOsupp}
  Y. Kanada-En'yo and H. Horiuchi,
  Prog. Theor. Phys. Suppl. {\bf 142}, 205 (2001).
\bibitem{AMDrev} 
  Y. Kanada-En'yo M. Kimura and H. Horiuchi, 
  C. R. Physique {\bf 4} 497 (2003).
%\bibitem{AMDrev2}
%  Y. Kanada-En'yo and M. Kimura,
%  lecture notes in physics {\bf 818}, 129 (2010).


\bibitem{ENYO-c12}
 Y. Kanada-En'yo,
Phys. Rev. Lett. {\bf 81}, 5291 (1998).
\bibitem{KanadaEn'yo:2006ze} 
  Y.~Kanada-En'yo,
  %``Structure of ground and excited states of C-12,''
  Prog.\ Theor.\ Phys.\  {\bf 117}, 655 (2007)
  [Erratum-ibid.\  {\bf 121}, 895 (2009)].
%  [nucl-th/0605047].
  %%CITATION = NUCL-TH/0605047;%%

\bibitem{Chernykh07} 
M.Chernykh, H.Feldmeier, T.Neff, P.von Neumann-Cosel and A.Richter, 
Phys. Rev. Lett. {\bf 98}, 032501 (2007).


\bibitem{brink66} D. M. Brink, International School of Physics ``Enrico Fermi'', XXXVI, p. 247 (1966).



%\bibitem{BCS}
%J.~Bardeen, L.~N.~Cooper and J.~R.~Schrieffer, Phys. Rev. {\bf 108}, 1175 (1957).  





\end{thebibliography}
\end{document}